\documentclass[a4paper,11pt]{article}
\pdfoutput=1 % if your are submitting a pdflatex (i.e. if you have
\usepackage{jheppub} % for details on the use of the package, please
\usepackage[T1]{fontenc}
\usepackage[italian,english]{babel}
\usepackage{hyperref}
\usepackage{ifpdf}
\usepackage{subfigure}
\usepackage{amssymb}
\usepackage{amsfonts}
\usepackage{epsf}
\usepackage{rotating}
\usepackage{graphicx}
\usepackage{amsmath}
\usepackage{fancyhdr}
\usepackage{lineno}
\usepackage{babel}
\usepackage{graphics}
\usepackage{pstricks}
\usepackage{color}
\usepackage{multirow}
\newcommand{\nn}{\nonumber}
\newcommand{\lsim}{\mathrel{\mathop{\kern 0pt \rlap
  {\raise.2ex\hbox{$<$}}}
  \lower.9ex\hbox{\kern-.190em $\sim$}}}
\newcommand{\gsim}{\mathrel{\mathop{\kern 0pt \rlap
  {\raise.2ex\hbox{$>$}}}
  \lower.9ex\hbox{\kern-.190em $\sim$}}}

\newcommand{\be}{\begin{equation}}
\newcommand{\ee}{\end{equation}}
\newcommand{\bea}{\begin{eqnarray}}
\newcommand{\eea}{\end{eqnarray}}

\def\ptmiss{\not\!\!{p_T}}
\title{\boldmath Probing the Hidden Higgs Bosons of the $Y=0$ Triplet- and Singlet-Extended Supersymmetric Standard Model at the LHC}

\author[a]{Priyotosh Bandyopadhyay}

\author[a,b]{Claudio Corian\`o}
\author[a]{Antonio Costantini}

\affiliation[a]{Dipartimento di Matematica e Fisica "Ennio De Giorgi", \\ Universit\`a del Salento and INFN-Lecce, \\ Via Arnesano, 73100 Lecce, Italy}

\affiliation[b]{STAG Research Centre and Mathematical Sciences,\\ University of Southampton, Southampton SO17 1BJ, UK}

\emailAdd{priyotosh.bandyopadhyay@le.infn.it}
\emailAdd{claudio.coriano@le.infn.it}
\emailAdd{antonio.costantini@le.infn.it}

\abstract{ We investigate the scalar sector in an extension of the Minimal
Supersymmetric Standard Model (MSSM) containing a $SU(2)$ Higgs triplet of zero hypercharge and a gauge singlet beside the $SU(2)$ scalar doublets. In particular, we focus on a scenario of this model which allows a light pseudoscalar and/or a scalar
below $100$ GeV, consistent with the most recent data from the LHC and the earlier data from the LEP experiments. We analyze the exotic decay of the discovered Higgs $(h_{125})$ into two light (hidden) Higgs bosons present in the extension. The latter are allowed by the uncertainties in the Higgs decay  $h_{125}\to WW^*$, $h_{125}\to ZZ^*$ and $h_{125}\to \gamma\gamma$. The study of the parameter space for such additional scalars/pseudoscalars decay of the Higgs is performed in the gluon fusion channel. The extra hidden Higgs bosons of the enlarged scalar sector, if they exist, will then decay into
lighter fermion paris, i.e., $b\bar{b}$, $\tau\bar{\tau}$ and $\mu\bar{\mu}$ via the mixing with the doublets. A detailed simulation using PYTHIA of the $2b+2\tau$, $\geq 3\tau$, $2b+2\mu$ and $2\tau+2\mu$ final states is presented. From our analysis we conclude that, depending on the selected benchmark points,
such decay modes can be explored with an integrated luminosity of 25 fb$^{-1}$ at the LHC at a center of mass energy of 13 TeV.}

\begin{document}

\maketitle
\flushbottom

\section{Introduction}
The success of the Standard Model (SM) in explaining the gauge structure of the fundamental interactions has reached its height with the discovery of a scalar particle with {most of} the properties of the SM Higgs boson - as a 125 GeV mass resonance - at the LHC. With this discovery, the mechanism of spontaneous symmetry breaking of the gauge symmetry, which in a gauge theory such as the SM is mediated by a Higgs doublet, has been confirmed, but the possible existence of an extended Higgs sector, at the moment, cannot be excluded. 

The identification by the CMS \cite{CMS, CMS2} and ATLAS \cite{ATLAS} experiments of a new boson exchange, has interested so far only the $WW^*$, $ZZ^*$ and $\gamma\gamma$
channels - using data at 7 and at 8 TeV - at more than $5\sigma$ confidence level for the $Z$ and $\gamma$ cases, and slightly below in the $W$ channel. However, the fermionic decay modes of the new boson, together with other exotic decay modes, are yet to be discovered. Clearly, they are essential in order 
to establish the mechanism of electroweak symmetry breaking (EWSB), which is crucial in the SM dynamics, with better precision.  
The new data collection at the LHC at 13 TeV center of mass energy - which will be upgraded to 14 TeV in the future - will probably provide new clues about some possible extensions of the SM, raising large expectations both at theoretical and at experimental level. \\
The SM is not a completely satisfactory theory, even with its tremendous success, since it does not provide an answer to long-standing issues, most prominently the gauge-hierarchy problem. This is instead achieved by the introduction of supersymmetry, which, among its benefits, allows gauge coupling unification and, in its R-parity conserving version,   
also provides a neutral particle as a dark matter candidate. The absence of any supersymmetric signal at the LHC and the recent observation of  a Higgs boson $(h_{125})$ of 125 GeV in mass, requires either a high SUSY mass scale or larger mixings between the scalar tops \cite{pMSSMb}. The situation is severer for more constrained 
SUSY scenarios like mSUGRA  \cite{cMSSMb}, which merge supersymmetric versions of the SM with minimal 
supergravity below the Planck scale. 

In the current situation, extensions of the Higgs sector with the inclusion of one or more electroweak doublets and/or of triplets of different hypercharges - in combination with {SM gauge} singlets - are still theoretical 
possibilities in both supersymmetric and non-supersymmetric extensions of the SM. We have recently shown that a supersymmetric extension of SM with a  $Y=0$ triplet and a singlet Higgs superfields \cite{TNSSMo}, called the TNMSSM, is still a viable scenario, which is compatible with the recent LHC results and the previous constraints from LEP, while respecting several others direct and indirect experimental limits. Building on our previous analysis, here we are going to show 
that the same model allows a light pseudoscalar in the spectrum, which could have been missed both by older searches at LEP \cite{LEPb} and by the recent ones at the LHC \cite{CMS, CMS2, ATLAS}.\\
Concerning the possible existence of an extended Higgs sector, the observation of a Higgs boson decaying into two light scalar or pseudoscalar states would be one of its direct manifestations. 
This detection would also allows us to gather significant information about the cubic couplings of the Higgs and, overall, about its potential. However, so far neither the CMS nor the ATLAS collaborations have presented direct bounds 
on the decays of the Higgs $h_{125}$ into two scalars.  If such scalars are very light ($m_\Phi\lsim 100$ GeV), then they cannot be part of the spectrum of an ordinary CP-conserving minimal supersymmetric extension of the SM (MSSM). In fact, in that case they are predicted to be accompanied by a heavy pseudoscalar or by a charged Higgs boson. The only possibilities which are left open require CP-violating scenarios where one can have a light scalar  with a mostly CP-odd component \cite{CPVMSSM}. Such scenarios, however, are in tension with the recent observations of the decay mode $h\to \tau \tau$ \cite{CPVMSSMb}.\\
 The natural possibilities for such hidden Higgs bosons are those scenarios characterized by an extended Higgs sector. In the next-to-minimal supersymmetric standard model (NMSSM) with a $Z_3$ symmetry, such a light
 pseudoscalar is part of the spectrum in the form of a pseudo Nambu-Goldstone mode \cite{NMSSMps}. This situation gets even more interesting with the addition of triplets of appropriate hypercharge assignments \cite{TNSSMo,tnssm}, as in the TNMSSM. In the case of  a $Y=0$ Higgs triplet- and singlet-extended scenarios, the triplet does not couple to the $Z$ boson and the singlet to any gauge boson, and both of them do not couple to fermions.\\
At  LEP the Higgs boson was searched in the mass range less than $114.5$ GeV via the production of $e^+e^-\to Zh$ and  $e^+e^-\to h_ia_j$
 (in scenarios with two Higgs doublets), involving scalar $(h_i)$ and pseudoscalar $(a_j)$ with  fermionic final states. The $Y=0$ TNMSSM thus becomes a natural candidate for the
 such hidden Higgs possibility and therefore can evade the LEP bounds \cite{LEPb}.
 However, the situation gets slightly more complicated for Higgs triplets of non-zero hypercharge because they do couple to the $Z$ boson. 
 
 %{\bf\blue The finding of such hidden scalars would certainly be a proof of the existence of an extended Higgs sector, 
% but the identification of such states as being of triplet/singlet origin
% would require more effort. We remark that the NMSSM does not have any extra charged Higgs boson compared to the MSSM, wheras the TNMSSM 
% includes one an extra triplet-like charged Higgs boson which does not couple to fermions and can decay to $h^\pm \to Z W^\pm$ \cite{tssmch1prime}.
% Therefore, the identification of Higgs scalars which are part of new representations, compared to the usual $SU(2)$ doublet representation, we need to explore the charged Higgs sectors more carefully. A possibility of 
% a very light triplet-like charged Higgs cannot be ruled out in the TNMSSM, which can still allow an exotic Higgs decay, $h_{125} \to h^\pm_i W^\mp$.}
 
 In this article we will focus our attention on decays of the Higgs boson into light scalars and pseudoscalars ($h_{125} \to h_ih_j/a_ia_j$). Such light scalar or pseudoscalars, when characterized by a mostly triplet or singlet component, do not 
 couple directly to fermions but decay to fermion pairs ($b$ or $\tau$) via their mixing with Higgs bosons of doublet type under $SU(2)$. Thus their final states 
 are often filled up with $b$-quarks, and leptons $\tau$ and $\mu$'s. The corresponding leptons and jets are expected to be rather soft, depending on the masses of the hidden scalars. If the doublet-triplet/singlet mixings in the Higgs sector are very small, they can give rise to the typical leptonic signature of charged displaced vertices. The goal of our analysis is to provide a direct characterization of the final states in the decay of a Higgs-like particle which can be helpful in the search for such hidden scalars at the LHC.
 
 It is organized as follows. After a brief overview of the TNMSSM in section~\ref{model}, we investigate in section~\ref{ggh} the decays of the Higgs to a gluon pair and calculate the decay to two pseudoscalars in section~\ref{psdcy}. In section~\ref{secbps}
 we discuss the phenomenology of the hidden Higgs bosons and select some benchmark points for a collider study at typical LHC energies. In 
 section ~\ref{sigsim} we perform a detail collider simulation for the signal and consider all the dominant SM backgrounds
  for the chosen final states, presenting the relative results, before our conclusions, which are contained in section~\ref{concl} .

\section{The Model}\label{model}
As detailed in \cite{TNSSMo},  the superpotential of the TNMSSM, $W_{TNMSSM}$, contains a SU(2) triplet $\hat{T}$ of zero hypercharge ($Y=0$)  together with a SM gauge singlet ${\hat S}$ added to the superpotential of the MSSM. Its structure can be decomposed in the form 
 \begin{equation}
 W_{TNMSSM}=W_{MSSM} + W_{TS},
 \end{equation}
with
\begin{equation}
W_{MSSM}= y_t \hat U \hat H_u\!\cdot\! \hat Q - y_b \hat D \hat H_d\!\cdot\! \hat Q - y_\tau \hat E \hat H_d\!\cdot\! \hat L\ ,
\label{spm}
 \end{equation}
 being the superpotential of the MSSM, while 
 \begin{equation}
W_{TS}=\lambda_T  \hat H_d \cdot \hat T  \hat H_u\, + \, \lambda_S \hat S  \hat H_d \cdot  \hat H_u\,+ \frac{\kappa}{3}\hat S^3\,+\,\lambda_{TS} \hat S  \textrm{Tr}[\hat T^2]
\label{spt}
 \end{equation}
accounts for the extended scalar sector which includes a spin triplet and a singlet superfields. In our notation a ''$\cdot$'' denotes a contraction with the Levi-Civita symbol $\epsilon^{ij}$, with $\epsilon^{12}=+1$  The triplet and doublet superfields are given by 

\begin{equation}\label{spf}
 \hat T = \begin{pmatrix}
       \sqrt{\frac{1}{2}}\hat T^0 & \hat T_2^+ \cr
      \hat T_1^- & -\sqrt{\frac{1}{2}}\hat T^0
       \end{pmatrix},\qquad \hat{H}_u= \begin{pmatrix}
      \hat H_u^+  \cr
       \hat H^0_u
       \end{pmatrix},\qquad \hat{H}_d= \begin{pmatrix}
      \hat H_d^0  \cr
       \hat H^-_d
       \end{pmatrix}.
 \end{equation}
 Here $\hat T^0$ is a complex neutral superfield, while  $\hat T_1^-$ and $\hat T_2^+$ are the charged Higgs superfields.  
 The MSSM Higgs doublets are the only superfields which couple to the fermion multiplet via Yukawa coupling as in Eq.~(\ref{spm}). The singlet and the triplet superfields account for the supersymmetric $\mu_D$ term coupling $H_u$ and $H_d$, after that their neutral components acquire vacuum expectation values in Eq.~(\ref{spt}).
 
 %\hat H_d \cdot  \hat H_u $ term after that ${\hat S}$ acquires a vev, with 
%$ \mu_D=\lambda_S v_S$, as shown in Eq.~(\ref{spt}).\\
It is a characteristic of any scale invariant supersymmetric theory with a cubic superpotential that the complete Lagrangian with the soft SUSY breaking terms has an accidental  $Z_3$ symmetry. This is generated by the invariance of all of its components 
after multiplication of the chiral superfields by the phase $e^{2\pi i/3}$ which, as we are going to discuss below, affects the mass of the pseudoscalars. 

The soft breaking terms in the scalar potential are given by

 \bea\nn
V_{soft}& =&m^2_{H_u}|H_u|^2\, +\, m^2_{H_d}|H_d|^2\, +\, m^2_{S}|S|^2\, +\, m^2_{T}|T|^2\,+\, m^2_{Q}|Q|^2 + m^2_{U}|U|^2\,+\,m^2_{D}|D|^2 \\ \nn
&&+(A_S S H_d.H_u\, +\, A_{\kappa} S^3\, +\, A_T H_d.T.H_u \, +\, A_{TS} S Tr(T^2)\\ 
 &&\,+\, A_U U H_U . Q\, +\, \, A_D D H_D . Q + h.c),
\label{softp}
 \eea
while the D-terms take the form 
 \begin{equation}
 V_D=\frac{1}{2}\sum_k g^2_k ({ \phi^\dagger_i t^a_{ij} \phi_j} )^2 .
 \label{dterm}
 \end{equation}
 As in our previous study, also in this case we assume that all the coefficients involved in the Higgs sector are real in order to preserve CP invariance. The breaking of the $SU(2)_L\times U(1)_Y$ electroweak symmetry is then obtained by giving real vevs to the neutral components of the Higgs field
 \be
 <H^0_u>=\frac{v_u}{\sqrt{2}}, \, \quad \, <H^0_d>=\frac{v_d}{\sqrt{2}}, \quad ,<S>=\frac{v_S}{\sqrt{2}} \, \quad\, <T^0>=\frac{v_T}{\sqrt{2}},
 \ee
 which give mass to the $W^\pm$ and $Z$ bosons
 \be
 m^2_W=\frac{1}{4}g^2_L(v^2 + 4v^2_T), \, \quad\ m^2_Z=\frac{1}{4}(g^2_L \, +\, g^2_Y)v^2, \, \quad v^2=(v^2_u\, +\, v^2_d), 
\quad\tan\beta=\frac{v_u}{v_d} \ee
 and also induce, as mentioned above, a $\mu$-term of the form $ \mu_D=\frac{\lambda_S}{\sqrt 2} v_S+ \frac{\lambda_T}{2} v_T$.
 The triplet vev $v_T$ is strongly  constrained by the global fit on the measurement of the $\rho$ parameter \cite{rho}
 \be
 \rho =1.0004^{+0.0003}_{-0.0004} ,
 \ee 
 which restricts its value to $v_T \leq 5$ GeV. Respect to the tree-level expression, the non-zero triplet contribution to the $W^\pm$ mass leads to a deviation of the $\rho$ parameter
 \be
 \rho= 1+ 4\frac{v^2_T}{v^2} .
 \ee
As in \cite{TNSSMo}, in our current numerical analysis we have chosen $v_T =3	$ GeV.

\section{Higgs decays into two gluons}\label{ggh}

\begin{figure}[t]
\begin{center}
\mbox{\hskip -20 pt\subfigure[]{\includegraphics[width=0.55\linewidth]{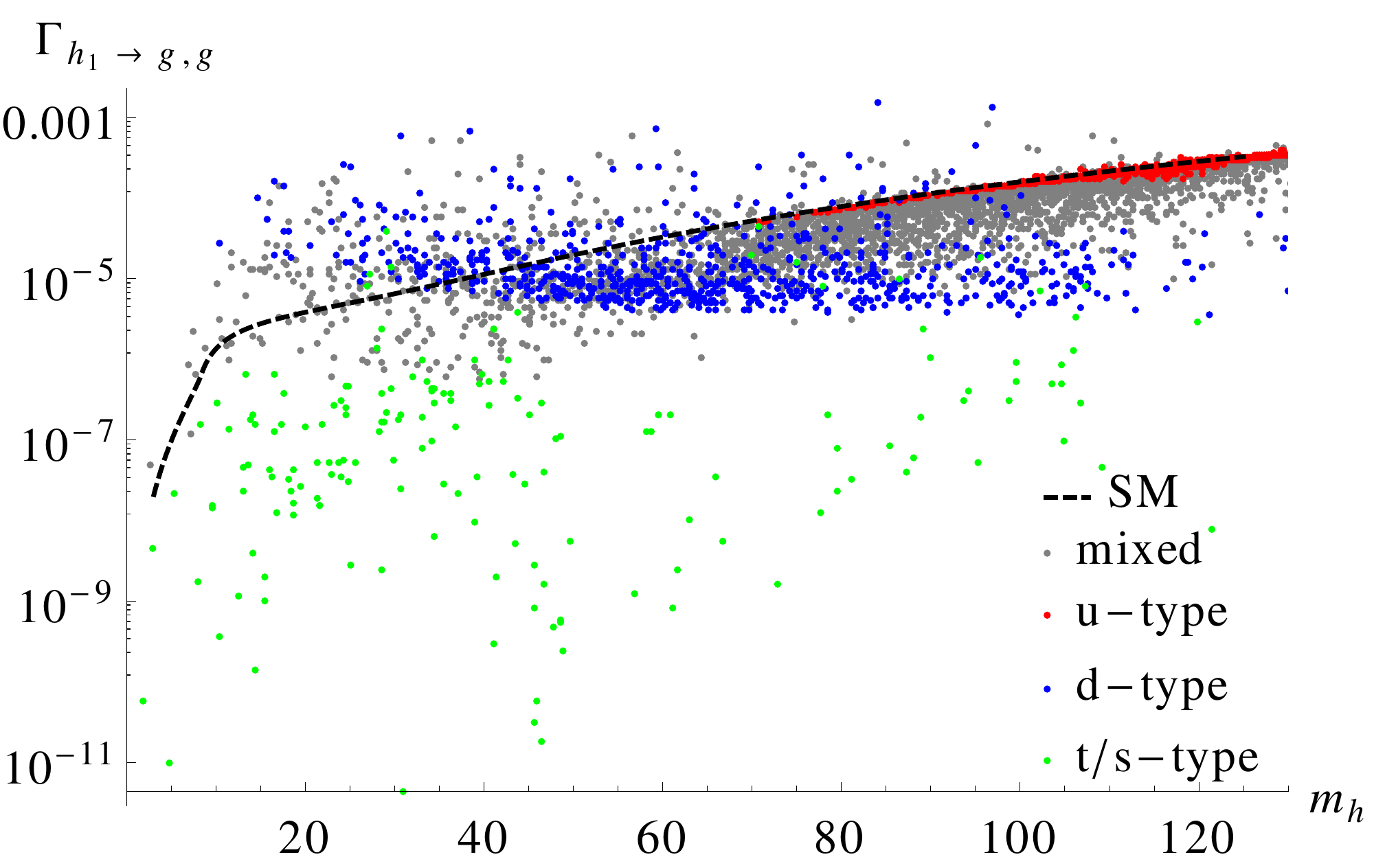}}
\subfigure[]{\includegraphics[width=0.55\linewidth]{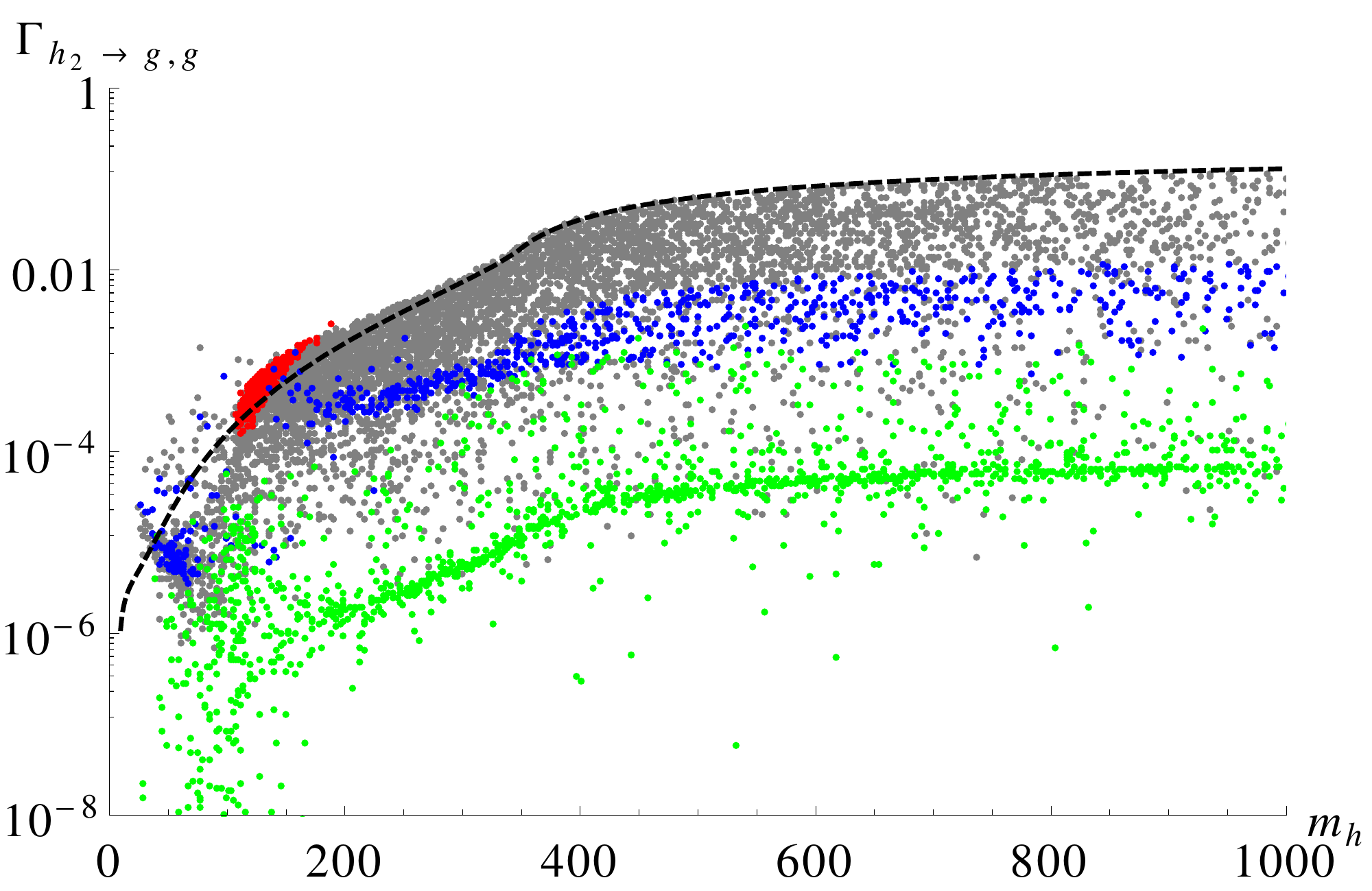}}}
\mbox{\hskip -20 pt\subfigure[]{\includegraphics[width=0.55\linewidth]{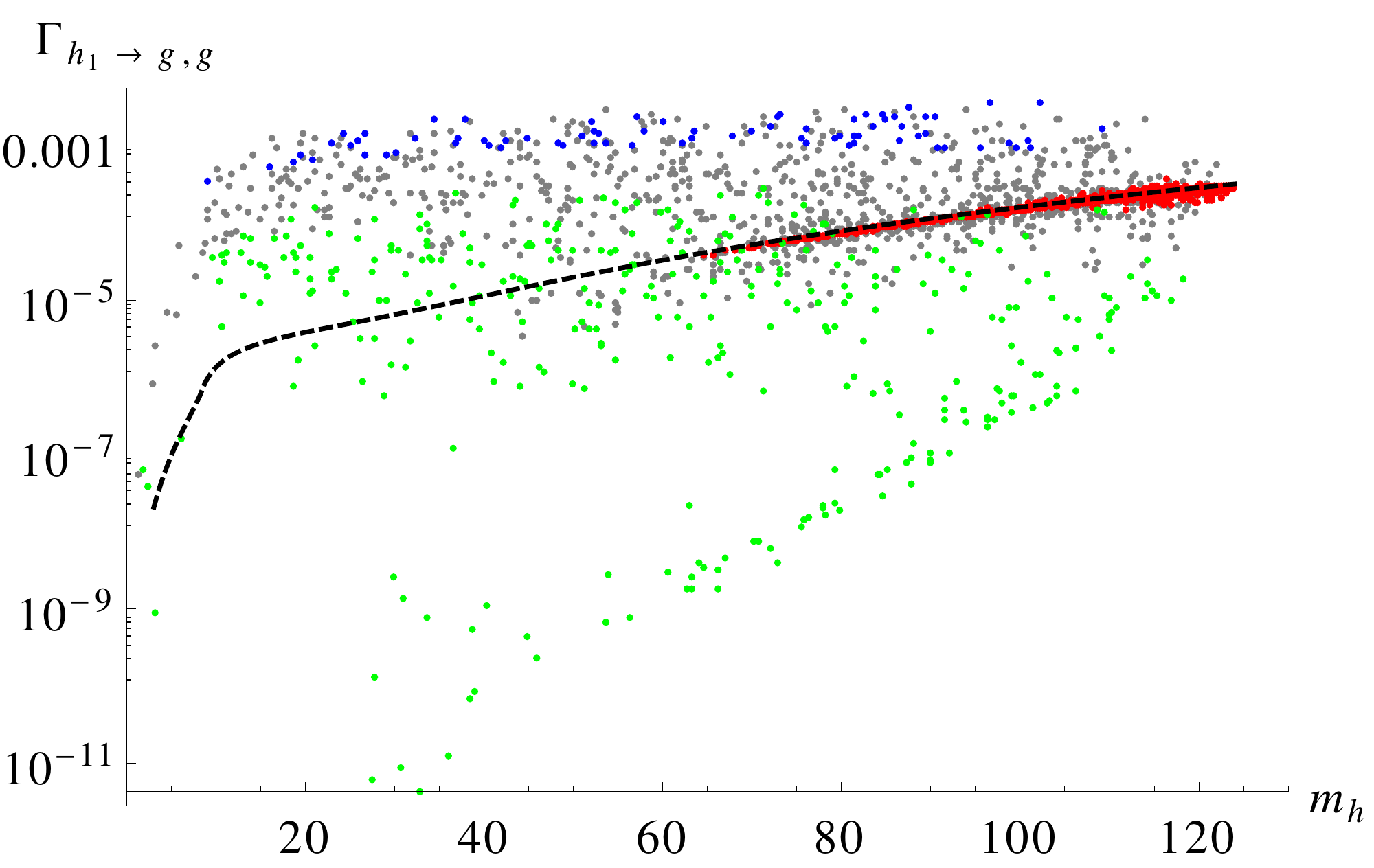}}
\subfigure[]{\includegraphics[width=0.55\linewidth]{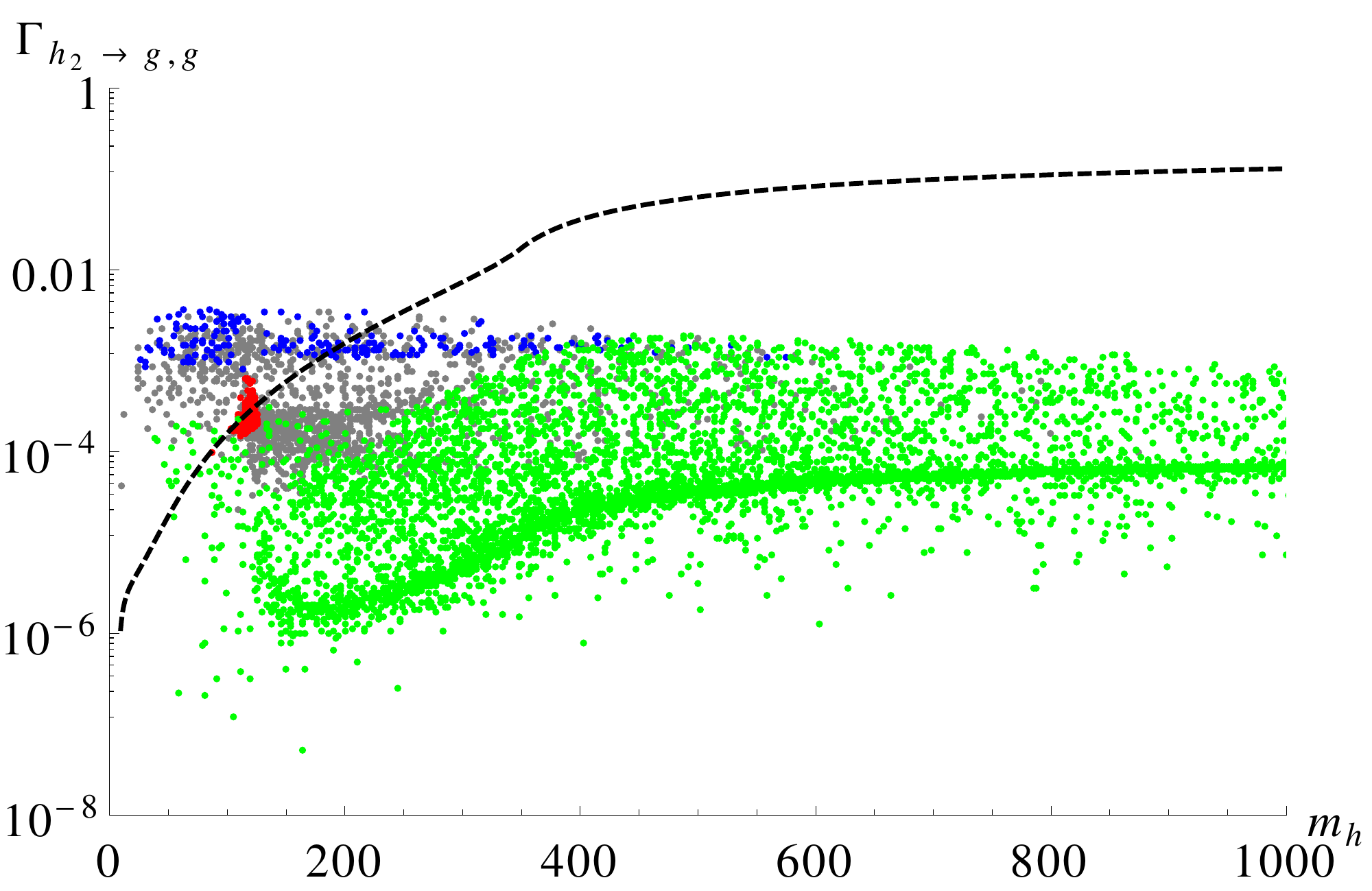}}}
\caption{We show a comparison between the SM and the TNMSSM predictions for the decay width of $h_1\rightarrow g,g$ (a), $h_2\rightarrow g,g$ (b) for $1<\tan\beta<15$ and $h_1\rightarrow g,g$ (c), $h_2\rightarrow g,g$ (d) for $20<\tan\beta<40$. We use the color code to distinguish among the up-type (>90\%) (red), down-type (blue), triplet/singlet-type (green) and mixed type Higgses (gray).}\label{hgg}
\end{center}
\end{figure}

 In the SM the most efficient production process of the Higgs boson is by gluon-gluon $(g)$ fusion (Figure (\ref{glutri})). The amplitude is mediated by a quark loop, which involves all the quarks of the SM, although the third generation, and in particular the top quark, gives the dominant contribution.
\begin{figure}[t]
\begin{center}
\includegraphics[width=0.3\linewidth]{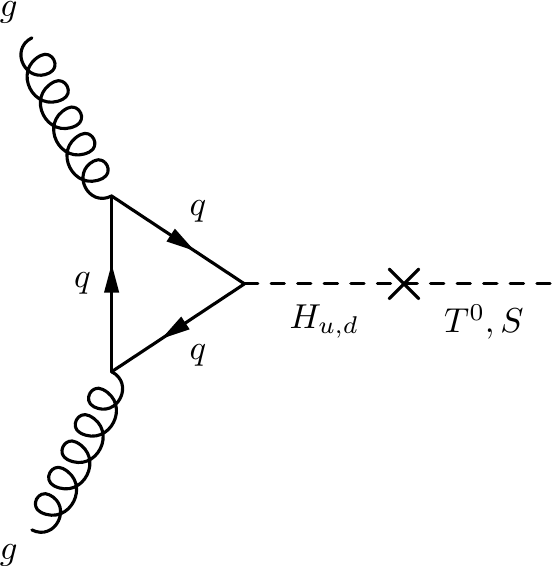}
\caption{A Feynman diagram depicting the coupling of gluons to the triplet/singlet, via their mixing with the doublets.}\label{glutri}
\end{center}
\end{figure}
In supersymmetric theories the situation is slightly different, because there are the up-type and down-type Higgs doublets 
$\hat H_u$ and $\hat H_d$  that couple to the up-type and down-type quarks/squarks respectively. Beside the sparticles contribution, the main difference between the SM and supersymmetric theories comes in the coupling of the Higgs bosons to fermions. These are given by
\bea
g_{h_i u\bar u} = -\frac{i}{\sqrt2}y_u \mathcal{R}^S_{i1},\\
g_{h_i d\bar d} = -\frac{i}{\sqrt2}y_d \mathcal{R}^S_{i2},\\
g_{h_i\ell\bar\ell} = -\frac{i}{\sqrt2}y_\ell \mathcal{R}^S_{i2},
\eea
where $R^S_{ij}$ is the rotation matrix of the CP-even sector. This means that the top/bottom contribution can be suppressed/enhanced, depending on the structure of $h_i$. The production cross section for $g,g\rightarrow h_i$ is related to the decay width of $h_i\rightarrow g,g$. At leading order, this decay width is given by
\begin{align}\label{gluongluon}
\Gamma(h_i\rightarrow g,g)&=\frac{G_F\,\alpha_s\,m_h^3}{36\sqrt2\,\pi^3} \left|\frac{3}{4}\sum_{q=t,\, b} \frac{g_{h_i q\bar q}}{(\sqrt2G_F)^{1/2}m_q}\, A_{1/2}(\tau^i_q)+\sum_{\tilde{q}=\tilde t, \, \tilde b}\frac{g_{h_i\tilde q\tilde q}}{m^2_{\tilde q}}A_0(\tau^i_{\tilde q})\right|^2,
\end{align}
where $A_0$ and $\,A_{1/2}$ are the spin-0 and spin-1/2 loop functions
\bea
&&A_0(x)=-\frac{1}{x^2}\left(x-f(x)\right),\\
&&A_{1/2}(x)=\frac{2}{x^2}\left(x+(x-1)f(x)\right),
\eea
with the analytic continuations 
\bea
f(x)=\left\{
\begin{array}{lr}
\arcsin^2(\sqrt{x})& x\leq1\\
-\frac{1}{4}\left(\ln\frac{1+\sqrt{1-1/x}}{1-\sqrt{1-1/x}}-i\pi\right)^2& x>1
\end{array}\right.
\eea
and $\tau^i_j=\frac{m_{h_i}^2}{4\,m_j^2}$. We show in Figure~\ref{hgg} the decay width of $h_{1,2}\rightarrow g,g$. In general, this decay width can be very different from the SM one in the case of supersymmetric theories with an extended Higgs sector, like the TNMSSM. In fact, in the latter case we have only the doublet Higgs that couples to the fermions, as shown in Eq. (\ref{spm}). This implies that if the Higgs is mostly triplet- or singlet-like, the fermion couplings are suppressed by $\mathcal{R}^S_{i1,2}$, in the limit of low $\tan\beta$. In Figure~\ref{hgg} the dashed line is the SM decay width and the color code is defined as follow:  we mark in red the up-type Higgs (>90\%), in blue the down-type, in green the triplet/singlet-type and in gray the mixed type. A look at Figure~\ref{hgg}(a) and (b) shows that for low $\tan\beta$ the decay width of a triplet/singlet-type Higgs is heavily suppressed. This occurs because the triplet and singlet Higgses couple to fermions only through the mixing with their analogue $SU(2)$ doublets. It is also rather evident that the shape of the decay widths for Higgses of up-type and of mixed-type are similar to those of the SM Higgs, for a large range of the mass of the extra Higgses. In Figure~\ref{hgg}(a) it is shown that for a light Higgs which takes the role of $h_{125}$, the SM decay width can be provided by the down-type Higgs of the TNMSSM, even in the case of low $\tan\beta$. Figure~\ref{hgg}(c) and (d) instead show that for a high value of $\tan\beta$ the decay width is dominated by the down-type Higgs, hence by the bottom quark. However it is still possible to have a SM-like decay width mediated by the top quark. In Figure~\ref{hgg}(d) it is quite evident that the bottom quark contribution has the same shape as in the MSSM \cite{anatomy}. In this case the TNMSSM decay width of the Higgs is very different from the SM one for $m_h\gsim200$ GeV.
 
\section{Higgs decays into pseudoscalars}\label{psdcy}

The most important consequence of the $Z_3$ symmetry of the potential is that the mass of the pseudoscalar is in the GeV range, $m_{a_1}\sim\mathcal O(10)$ GeV, if we choose $A_{S, T, TS, \kappa, U, D}\sim\mathcal O(1)$ GeV. In this situation the decay $h_{125}\rightarrow a_1,a_1$ can be kinematically allowed. We study the decay of $h_{125}\rightarrow a_1,a_1$ via the decay width, given by

\bea\label{haaWidth}
\Gamma_{h_i\rightarrow a_j,a_j}=\frac{G_F}{16\sqrt2\pi}\frac{M_Z^4}{M_{h_i}}\left(1-\frac{4\,M_{a_j}^2}{M_{h_i}^2}\right)\left|\frac{g_{h_ia_ja_j}}{i M_Z^2/v}\right|^2,
\eea
where the $g_{h_ia_ja_j}$ coupling is given in the appendix. In Figure~\ref{Whaa}(a) and (b) we plot this decay width as a function of $\lambda_S$ and $\lambda_T$ respectively. Figure~\ref{Whaa}(a) shows that for $\left|\lambda_S\right|\gsim0.3$ we have scenarios in which the Higgs of doublet-type decays into pseudoscalars of singlet-type, but Figure~\ref{Whaa}(b) shows no particular structure in the dependence of $\Gamma_{h_1\rightarrow a_1,a_1}$ on $\lambda_T$.

Being interested in the fermionic final states of the decay of the SM-like Higgs into the light pseudoscalar $a_1$, $h_{125}\rightarrow a_1,a_1$, we gather the relevant coupling of the same pseudoscalars to fermions, which are given by  
\bea
g_{a_i u\bar u} = -\frac{\gamma_5}{\sqrt2}y_u \mathcal{R}^P_{i1},\\
g_{a_i d\bar d} = -\frac{\gamma_5}{\sqrt2}y_d \mathcal{R}^P_{i2},\\
g_{a_i\ell\bar\ell} = -\frac{\gamma_5}{\sqrt2}y_\ell \mathcal{R}^P_{i2}.
\eea 
Because the triplet, as well as the singlet, do not couple to the fermions, each $a_i$ will decay into fermions only trough a mixing with the doublet Higgses. This means that if $a_1$ is mostly of triplet or singlet component, its fermionic decay will be suppressed by the rotation elements $\mathcal{R}^P_{i1,2}$. An interesting consequence of this property is that this highly suppressed decay can generate a displaced vertex for the fermionic final states. 
\begin{figure}[t]
\begin{center}
\mbox{\hskip -20 pt\subfigure[]{\includegraphics[width=0.55\linewidth]{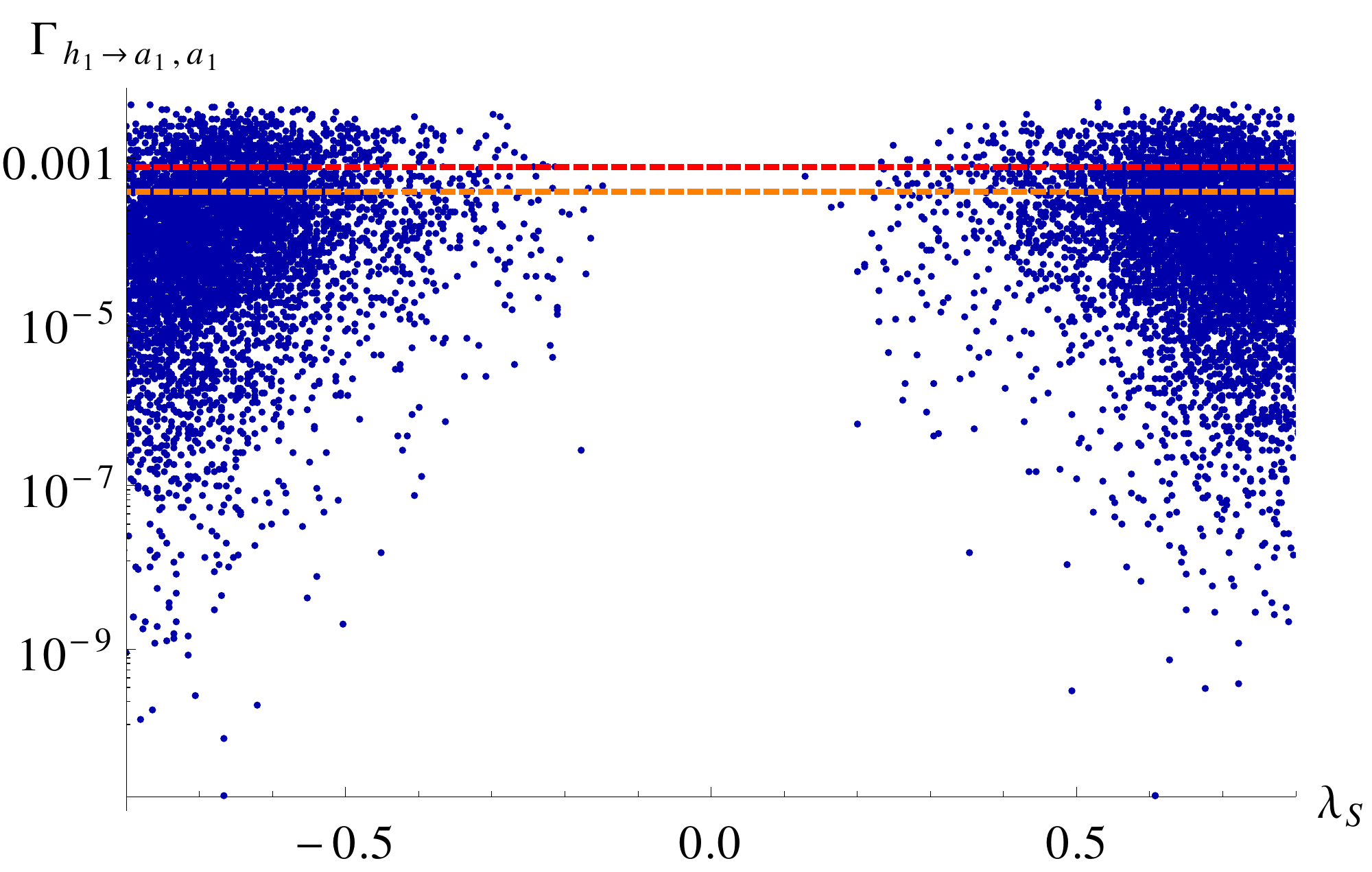}}
\subfigure[]{\includegraphics[width=0.55\linewidth]{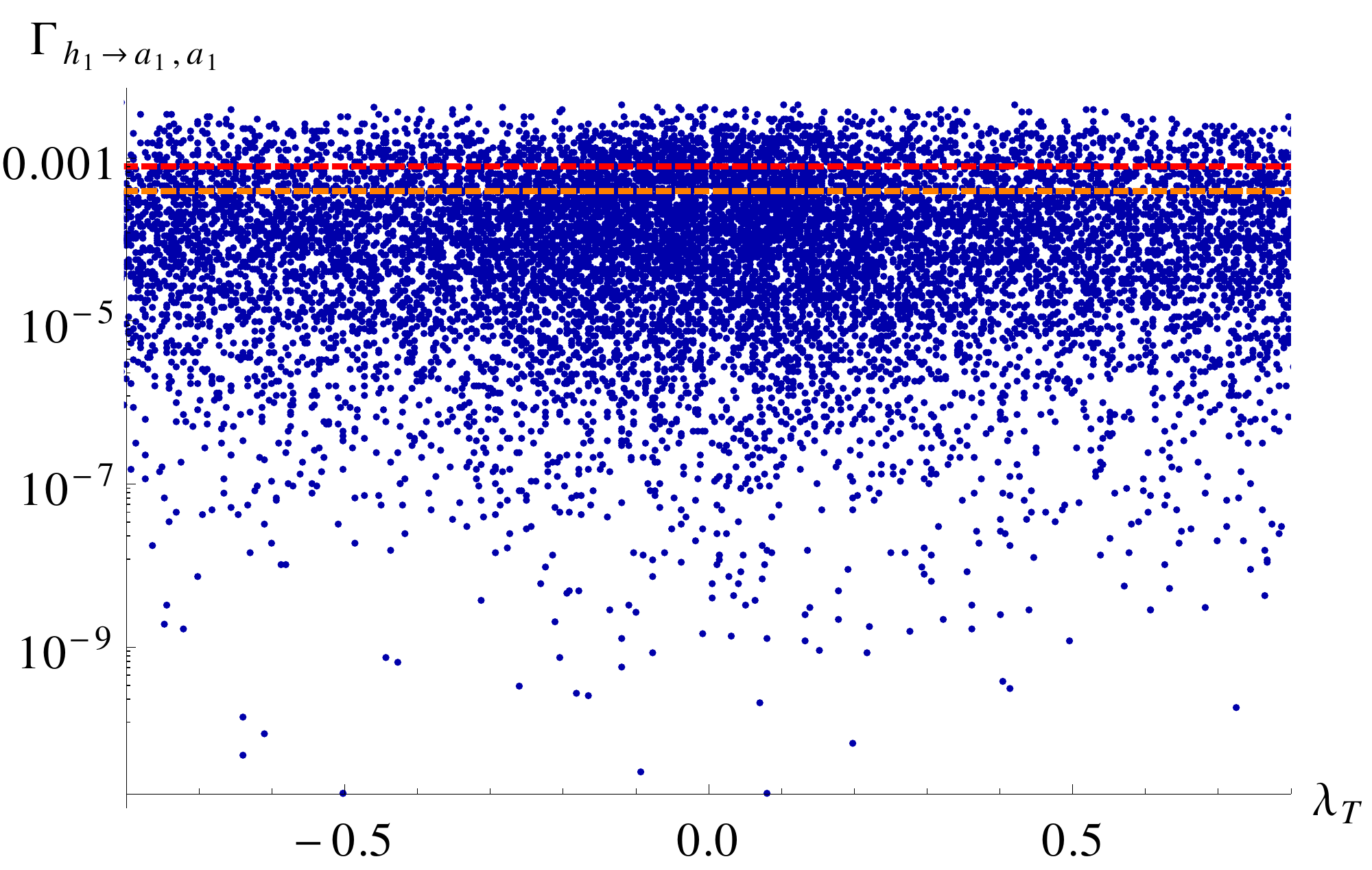}}}
\caption{We plot the decay width of the $h_{125}$ to two pseudoscalars (a) with respect to $\lambda_S$ and (b) with respect to $\lambda_{T}$. The red and orange coloured bands show the  region where $\mathcal{B}(h_{125} \to a_1 a_1)=20\%\, , 10\%$ respectively. }\label{Whaa}
\end{center}
\end{figure}

\section{Phenomenology and benchmark points}\label{secbps}

In Table~\ref{bps} we show the mass spectrum along with the other parameters which are necessary for the identification of three benchmark points. Together with the recent Higgs data we have also considered the recent bounds on the stop and sbottom masses \cite{thridgensusy} and the mass bounds on the lightest chargino from LEP \cite{chargino}. We have also taken into account the recent bounds on the charged Higgs boson mass from both CMS \cite{ChCMS} and ATLAS \cite{ChATLAS}. These have been derived in their searches for light in mass, charged Higgs bosons from the decay of a top quark, and in decays to $\tau \bar{\nu}$. The benchmark points 1 and 2 (BP1 and BP2) are characterized by one hidden Higgs boson, corresponding to a pseudoscalar particle of singlet-type with a mass of $\sim 20$ and $57$ GeV respectively. However BP3 has two hidden Higgs bosons, one of them a pseudoscalar of 
singlet-type around $\sim 37$ GeV and a second (scalar) one of triplet-type, around $\sim 118$ GeV in mass.
In the cases of BP1 \& BP2, $h_1$ is the discovered Higgs boson $h_{125}$, whereas for BP3 it is $h_2$.

%%%%%%%%%%%%%%%%%%%%%%%h_125 decays to a_1 a_1 %%%%%%%%%%%%%%
\begin{figure}[hbt]
\begin{center}
\mbox{\subfigure[]{\includegraphics[width=0.3\linewidth]{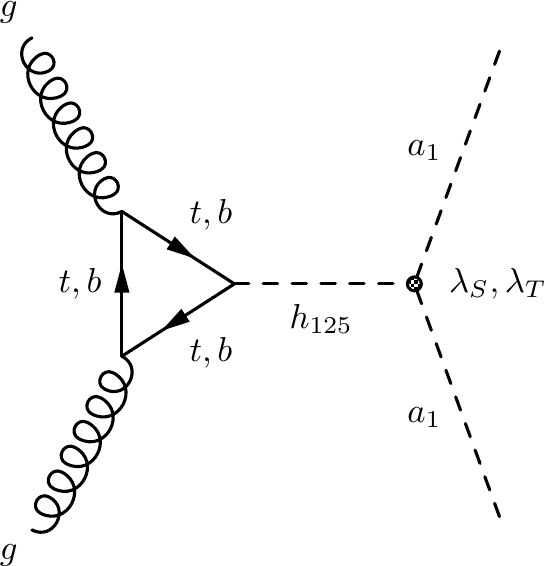}}\hskip 30 pt
\subfigure[]{\includegraphics[width=0.3\linewidth]{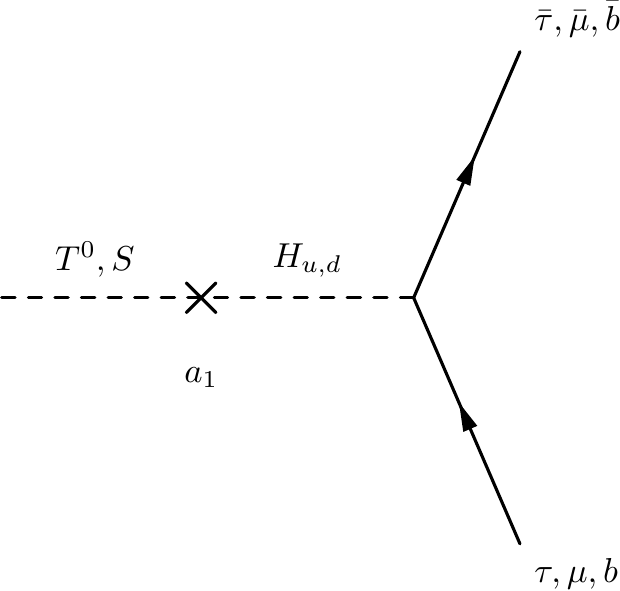}}}
\mbox{\subfigure[]{\includegraphics[width=0.4\linewidth]{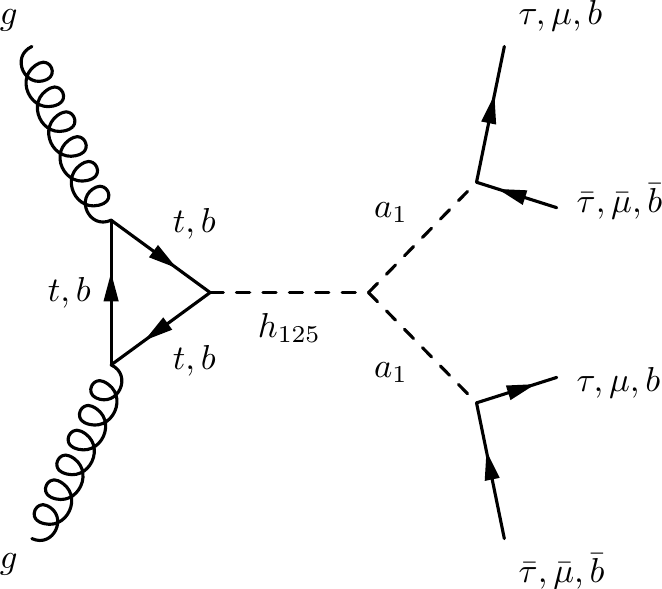}}}
\caption{Pseudoscalar (triplet/singlet) pair production from Higgs boson produced via 
gluon-gluon fusion and their decays, via their mixing with the doublets.}\label{glutri}
\end{center}
\end{figure}
 %%%%%%%%%%%%%%%%% Benchmark points%%%%%%%%%%%%%%%%%%%%
\begin{table}
\begin{center}
\renewcommand{\arraystretch}{1.4}
\begin{tabular}{||c||c|c|c||}
\hline\hline
Benchmark&BP1&BP2&BP3 \\
Points & &&\\ \hline\hline
$m_{h_1}$ & {\color{red}$\sim 125$} & {\color{red}$\sim 125$} & {\color{blue}$117.73 $} \\
\hline
$m_{h_2}$ & 183.58  &$162.59 $ & {\color{red}$\sim 125$} \\
\hline
$m_{h_3}$& 614.14 &$982.59$  & $791.37$ \\
\hline
$m_{h_4}$ & 965.75 &$1560.7$ & $1051.6$  \\
\hline
\hline
$m_{a_1}$ & \color{blue}20.50& \color{blue}$57.02$&  \color{blue}$36.79$ \\
\hline
$m_{a_2}$ &435.83  &$644.50$ & $620.81$  \\
\hline
$m_{a_3}$& 659.20 &$1018.1$  & $831.51$  \\
\hline
\hline
$m_{h^\pm_1}$ & \color{blue}182.84 &\color{blue}$162.25$ &\color{blue}$117.47$ \\
\hline
$m_{h^\pm_2}$ & 436.04 & $644.55$ & $620.86$  \\
\hline
$m_{h^\pm_3}$& 626.23 & $989.77$ & $805.58$   \\
\hline
\hline
$m_{\tilde{t}_1}$ &894.59 &$515.27$ & $460.47$    \\
\hline
$m_{\tilde{t}_2}$ & 961.10 &$835.45$ & $692.57$  \\
\hline
$m_{\tilde{b}_1}$& 629.08 & $491.37$ & $508.81$  \\
\hline
$m_{\tilde{b}_2}$& 948.54  &$790.93$  & $673.97$  \\
\hline
\hline
$\tan{\beta}$& 6.48 & $4.17$ & $3.55$    \\
\hline
\hline
\end{tabular}
\caption{Benchmark points for a collider study consistent with the $\sim 125$ GeV Higgs mass, where the $h_{i=1,2,3,4}$, $a_{i=1,2,3}$ are at one-loop and $h^{\pm}_{i=1,2,3}$ masses are calculated at tree level. We color in red the states which are mostly doublets ($>90\%$) and in blue those which are mostly triplet/singlet ($>90\%$). The points are consistent with the $2\sigma$ limits of $h_{125}\to WW^*, ZZ^*, \gamma\gamma$ \cite{CMS, ATLAS}.}\label{bps}
\end{center}
\end{table}
%%%%%%%%%%%%  
We now turn our attention to the decay of the discovered Higgs boson $h_{125}$ into a light pseudoscalar pair $a_1a_1$. Table~\ref{hdcy2} shows the branching ratios for the decay of $h_{125}$, in the case of the three benchmark points that we have selected. The table shows that for BP1 such branching ratio $(\mathcal{B})$ is the lowest $\mathcal{B}(h_{125}\to a_1a_1)\sim 10\%$, while for BP3 it is the highest $\mathcal{B}(h_{125}\to a_1a_1)\sim 18\%$. The discovered decay modes are consistent with the $2\sigma$ 
limits of $h_{125}\to WW^*, ZZ^*, \gamma\gamma$ \cite{CMS, ATLAS}. Such light pseudoscalars - though mostly singlet or
triplet - decay to the fermionic pairs which are kinematically allowed, via the mixing with the $H_u$ and $H_d$ doublets. This is because both singlet and triplet Higgses do not couples to fermions (see Eq.~\ref{spt}). 

For the benchmark point BP3 there is another hidden scalar which is CP-even, with a mass around $\sim 118$ GeV. $h_{125}$ cannot decay into this state $h_1$, as it is kinematically forbidden. If this $h_1$ is produced by other means it can have two-body decays to fermion pairs, as in the case of the $a_1$, via the mixing with the doublets. It will also have three-body decays ($WW^*$, $ZZ^*$) via its $SU(2)$ triplet charge and the mixing with the doublets.

%%%%%%%%%%%%%%%%% h_125 decay branching fraction (with tree level mass)%%%%%%%%%%%%%%%%%%%%
\begin{table}
\begin{center}
\renewcommand{\arraystretch}{1.4}
\begin{tabular}{||c||c|c|c|c|c|c|c||}
\hline\hline
Benchmark&\multicolumn{7}{|c||}{Branching ratios}\\
\hline
Points & $a_1 a_1$& $h_1 h_1$ & $a_1$Z &\; $W^+ W^-$ \;  & \;$b\bar{b}$ \;&\;$\tau \bar{\tau}$&$\mu\bar\mu$ \\
\hline\hline
BP1 &0.106  & - &$4.02\times10^{-7}$  & 0.138 &0.695  & 0.042&$1.50\times10^{-4}$\\
\hline
BP2 & 0.162 & - & $1.43\times10^{-8}$ & 0.136 & $0.645$ &$0.039$&$1.39\times10^{-4}$ \\
\hline
BP3 & $0.178$ & - & $1.93\times10^{-6}$ & 0.137 & $0.628$ & $0.038$&$1.35\times10^{-4}$ \\
\hline\hline
\end{tabular}

\caption{Decay branching ratios of $h_{125}$ for the three benchmark points, where the $h_{125}$ mass is calculated at tree level.
The kinematically forbidden decays are marked with dashes. The points are consistent with the $2\sigma$ limits of $h_{125}\to WW^*, ZZ^*, \gamma\gamma$ \cite{CMS, ATLAS}.}\label{hdcy2}
\end{center}
\end{table}
%%%%%%%%%%%%

%%%%%%%%%%%%%%%%% a_1 decay branching fraction (with tree level mass)%%%%%%%%%%%%%%%%%%%%
\begin{table}
\begin{center}
\renewcommand{\arraystretch}{1.4}
\begin{tabular}{||c||c|c|c||}
\hline\hline
Benchmark&\multicolumn{3}{|c||}{Branching ratios(\%)}\\
\hline
Points& \;$b\bar{b}$ \;&\;$\tau \bar{\tau}$&$\mu\bar\mu$  \\
\hline\hline
BP1 &   $0.939$ & $0.061$&$2.20\times10^{-4}$ \\
\hline
BP2 & $0.943$ & $0.057$&$2.04\times10^{-4}$\\
\hline
BP3 & $0.942$ & $0.058$& $2.07\times10^{-4}$\\
\hline\hline
\end{tabular}

\caption{Decay branching ratios of $a_1$ for the three benchmark points $BP_i$. The kinematically forbidden decays are marked with dashes.}\label{a1dcy2}
\end{center}
\end{table}
%%%%%%%%%%%%

For these benchmark points we have computed the production cross-sections of a $h_{125}$ Higgs boson assuming that  it is mediated by the gluon-gluon fusion channel
at the LHC. Table~\ref{Hcrosssec} presents the cross-sections which include the associated K-factors from the Higgs-Cross-Section
Working Group \cite{HCWG}. In the next section we are going to simulate the production of such light pseudoscalars
produced from the decay of such $h_{125}$. The choice of this particular production process is motivated by its large cross-section and by the rather clean final states ensued, that favour the extraction of the pseudoscalar $a_1$ pair.

%%%%%%%%%%%%%%%%% h_125 Gev Higgs cross-section at 13 TeV and 14 TeV %%%%%%%%%%%%%%%%

%%%%
\begin{table}
\begin{center}
\renewcommand{\arraystretch}{1.4}
\begin{tabular}{||c||c|c|c||}
\hline\hline
ECM&\multicolumn{3}{|c||}{$\sigma(gg\to h_{125}$) in pb}\\
in TeV&\multicolumn{3}{|c||}{for benchmark points}\\
\hline
&BP1 & BP2 &BP3 \\
\hline
13&41.00&41.00&41.00\\
\hline
14&46.18&46.18&46.18\\
\hline
\hline

\end{tabular}
\caption{Cross-section of $gg\to h_{125}$ at the LHC for center of mass energy of 13 and 14 TeV for the three benchmark points.}\label{Hcrosssec}
\end{center}
\end{table}

%%%%%%%%%%%%%%%%%%%%%%%%%%%%%%%%%%

\section{Signature and collider simulation}\label{sigsim}

The discovered Higgs boson $h_{125}$ can decay into two light pseudoscalars, which further decay into $\tau$ or $b$ pairs.  The $b$'s and $\tau$'s channel are therefore the relevant ones to look into, in the search for such hidden decay.  For this purpose we have implemented the model in SARAH \cite{sarah} and we have generated the model files for CalcHEP \cite{calchep}. These have been used to generate the decay file SLHA, containing the decay branching ratios and the corresponding mass spectra. The generated events have then been simulated with {\tt PYTHIA} \cite{pythia} via the the SLHA interface \cite{slha}. The simulation at hadronic level has been performed using the {\tt Fastjet-3.0.3} \cite{fastjet} with the {\tt CAMBRIDGE AACHEN} algorithm. We have selected a jet size $R=0.5$ for the jet formation, with the following criteria:
\begin{itemize}
  \item the calorimeter coverage is $\rm |\eta| < 4.5$

  \item the minimum transverse momentum of the jet $ p_{T,min}^{jet} = 10$ GeV and jets are ordered in $p_{T}$
  \item leptons ($\rm \ell=e,~\mu$) are selected with
        $p_T \ge 10$ GeV and $\rm |\eta| \le 2.5$
  \item no jet should be accompanied by a hard lepton in the event
   \item $\Delta R_{lj}\geq 0.4$ and $\Delta R_{ll}\geq 0.2$
  \item Since an efficient identification of the leptons is crucial for our study, we additionally require  
a hadronic activity within a cone of $\Delta R = 0.3$ between two isolated leptons to be $\leq 0.15\, p^{\ell}_T$ GeV, with 
$p^{\ell}_T$ the transverse momentum of the lepton, in the specified cone.

\end{itemize}

We keep the cuts in $p_T$ of the leptons and  the jets relatively low ($p_T \ge 10$ GeV), as they will be generated from the lighter pseudoscalar decays. $h_{125}$, once produced via  gluon-gluon fusion, will decay into two very light pseudoscalars ($m_{a_1}\sim 20$ GeV for BP1). 
The light pseudoscalars then will decay further into $b$ or $\tau$ pairs (see Table~\ref{a1dcy2}). The parton level signatures would be $4b$, $4\tau$ and $2b+2\tau$. 
In reality, this description is expected to change due to hadronization and to the contributions from the initial- and final-state 
radiation emission in the presence of $b$ quarks and of $\tau$ leptons. The number of jets can indeed
 increase or decrease due to these effects. The efficiency of the jet of the b-quark ($b_{\rm{jet}}$) is determined through the determination of the secondary vertex  and it is therefore momentum dependent. For this purpose
 we have taken - for the $b_{\rm{jet}}$'s from $t\bar{t}$ - the single-jet tagging efficiency equal to $0.5$, while for the remaining components of the final state we have followed closely the treatment of \cite{btag}. 
   Here, in the case of the $\tau_{\rm{jet}}$ we have considered the hadronic decay of the $\tau$ to be characterized by at least one charged track with $\Delta R \leq 0.1$ of the
  candidate $\tau_{\rm{jet}}$ \cite{taujet}.
%%%%%%%%%%%%%%%%%%%%%%%%%%%
\begin{figure}[hbt]
\begin{center}

\includegraphics[width=0.33\linewidth, angle=-90]{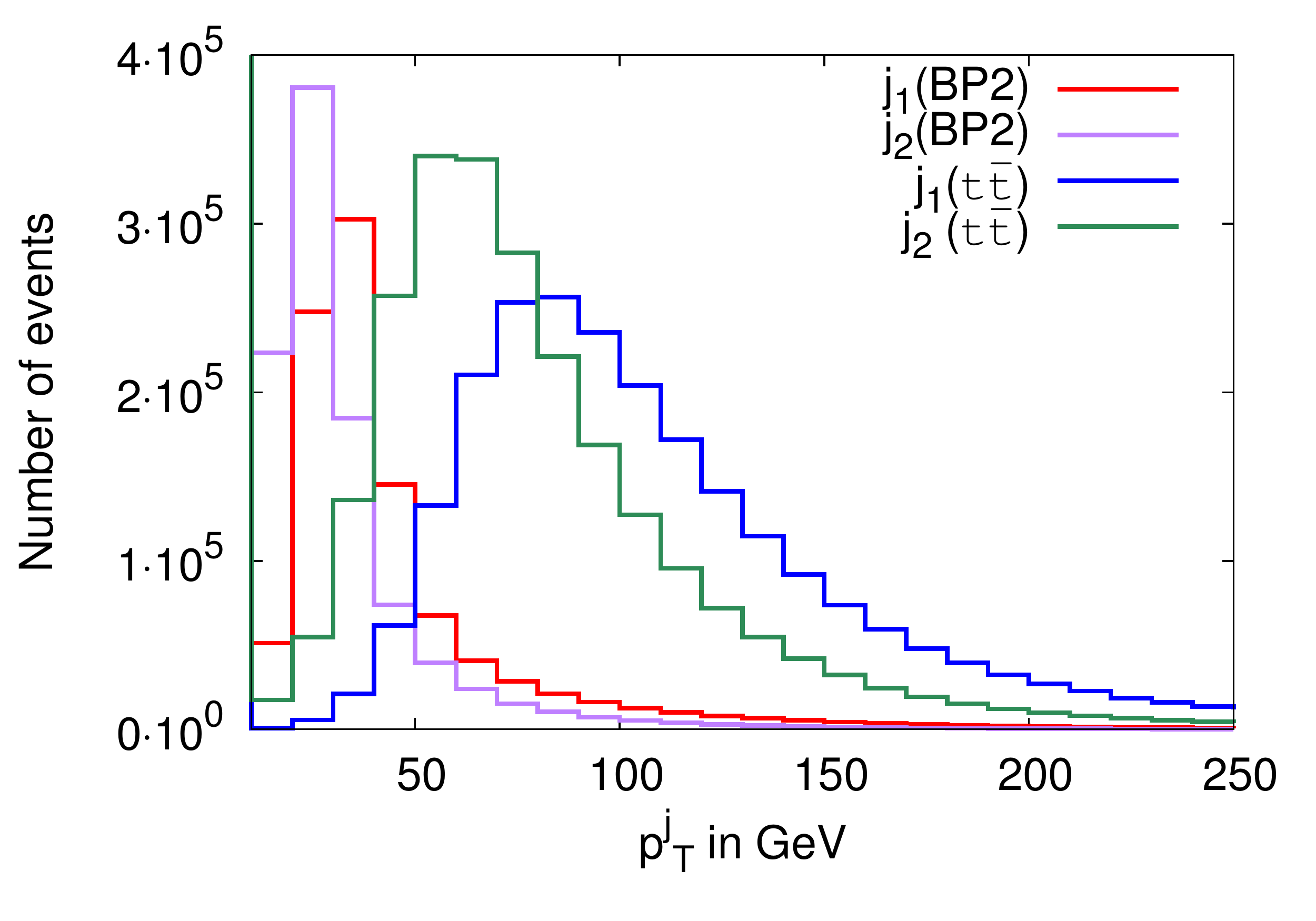}
\includegraphics[width=0.33\linewidth, angle=-90]{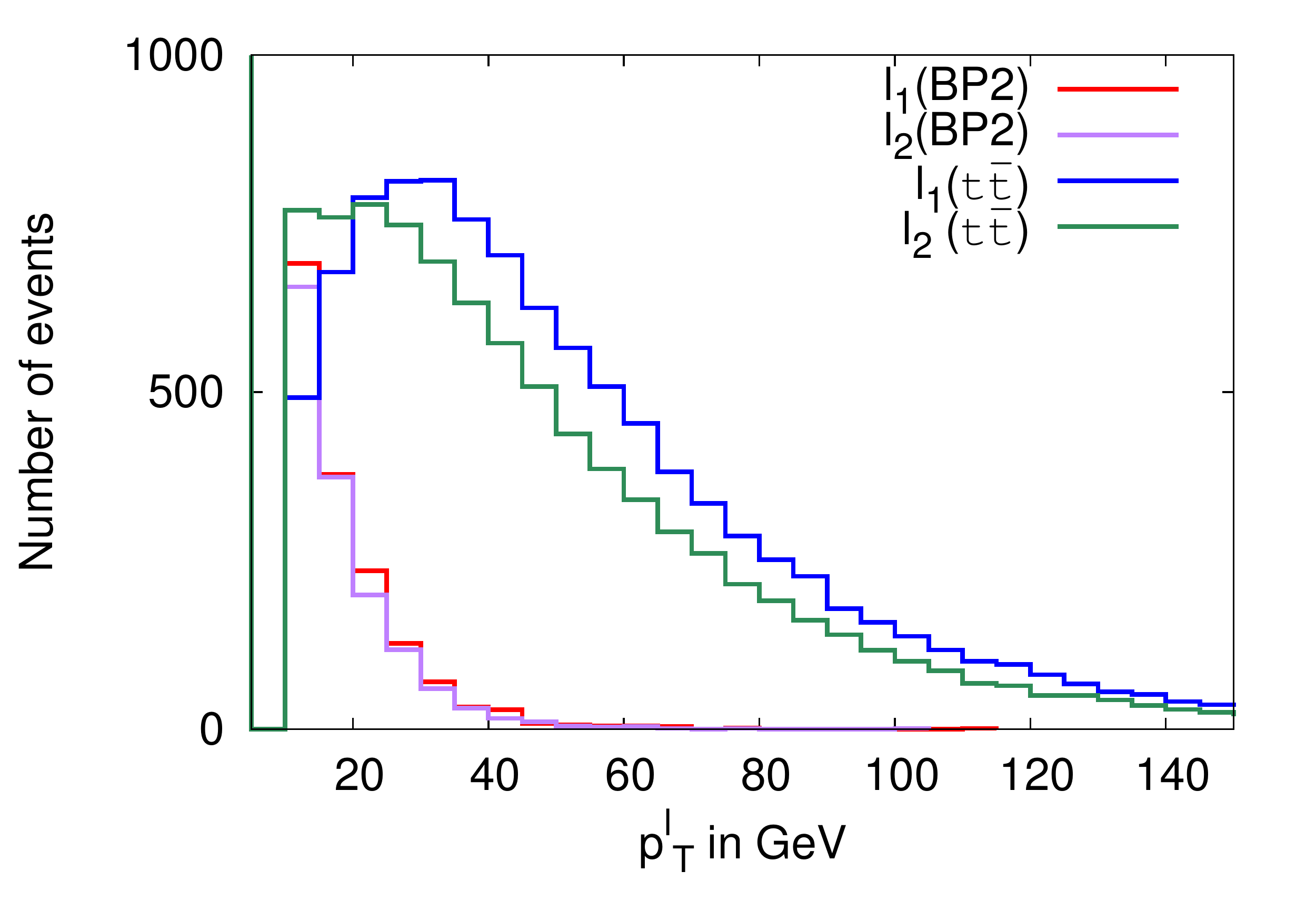}
\caption{ $p^{b_j}_T$ distribution (left) and  $p^{\ell}_T$ distribution (right) for $t\bar{t}$ and for the signal in BP2.}\label{ptjl}

\end{center}
\end{figure}
%%%%%%%%%%%%%%%%%%%%%%%%%%%%%%%%%%%%%%%%%%%%%%%%%%%%%%%%%%%%%%

Figure~\ref{ptjl} (left) shows the $b_{\rm{jet}}$  $p_T$ coming from the pseudoscalar decays in the case of BP2 with the dominant background $t\bar{t}$. 
Clearly one may observe the that $b_{\rm{jet}}$'s coming from the signal (BP2) are rather soft, mostly with $p_T \lesssim 50$ GeV. Figure~\ref{ptjl} (right) shows
the transverse momentum $p_T$ of the lepton coming from the signal (BP2) and the dominant backgrounds $t\bar{t}$ and $ZZ$. This clearly shows that the signal leptons 
are very soft ($p_T \lesssim 40$ GeV) compared to the corresponding backgrounds.

%%%%%%%%%%%%%%%%%%%%%%%%%%%
\begin{figure}[hbt]
\begin{center}

\includegraphics[width=0.33\linewidth, angle=-90]{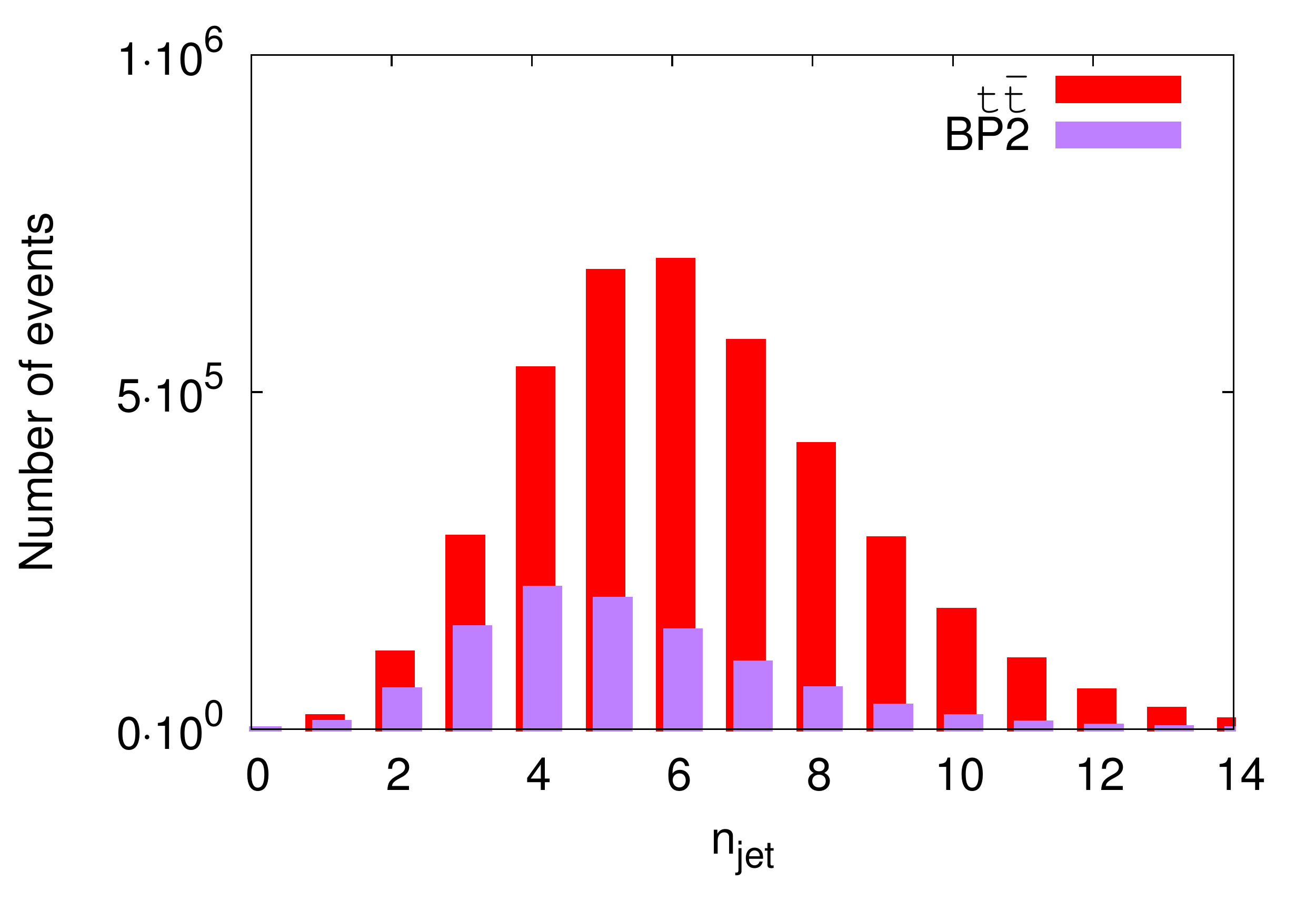}
\includegraphics[width=0.33\linewidth, angle=-90]{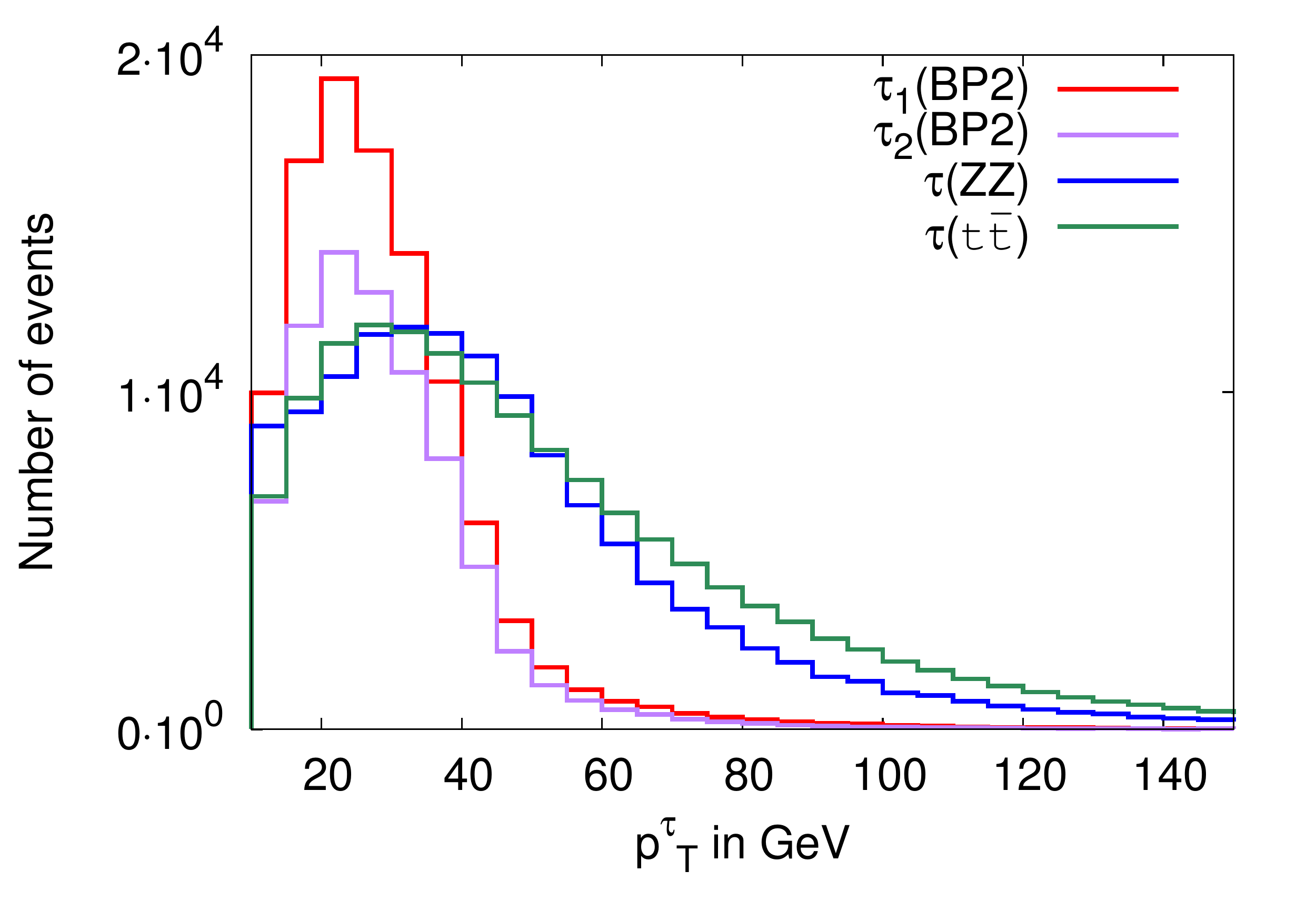}
\caption{(Left) jet-multiplicity ($n_{\textrm{jet}}$) distributions  and (Right) $p^{\tau}_T$ distributions for signal events coming from the  pseudoscalars $a_1$ decays for BP2 and the dominant SM backgrounds $t\bar{t}$, $ZZ$.}\label{njtaupt}

\end{center}
\end{figure}
%%%%%%%%%%%%%%%%%%%%%%%%%%%%%%%%%%%%%%%%%%%%%%%%%%%%%%%%%%%%%%
Next we have investigated the number of jets in the final states after hadronization. Figure~\ref{njtaupt} (left) 
shows the number of jets for the signal (BP2) and for the dominant background $t\bar{t}$. Due to the lower cuts in $p_T$, the number of final state jets has increased, in this case, both for the signal and for the background. The difference is still prominent between the two, where the signal peaks around 4 jets  and $t\bar{t}$ around 6. Thus a requirement of a relatively lower number of jets in the final state will remove the dominant $t\bar{t}$ contribution quite effectively.  

Figure~\ref{njtaupt} (right) shows the transverse momentum ($p^{\tau}_T$) distribution of the $\tau$ at parton level for the signal in BP2 and the dominant $\tau\tau$ backgrounds 
coming from $ZZ$ and $t\bar{t}$. Clearly, the condition of $p^{\tau}_T \lesssim 50$ GeV will reduce effectively the background contributions to the final state.  

\subsection{$2b+2\tau$}
%As we have already mentioned, the dominant production mode of the Higgs boson is the gluon-gluon fusion channel, 
%which has brought to the discovery of the $h_{125}$ state via three different channels by now; viz, $\gamma\gamma$, $ZZ^*$ and $WW^*$ by both CMS and ATLAS \cite{CMS, CMS2, ATLAS}. 
In the case of the TNMSSM, the discovered Higgs boson can also decay into a pair of  lighter mass eigenstates $a_1a_1$ and/or  $h_1h_1$. 
The possibility of producing such light states specially as singlet-like pseudoscalars has been discussed in \cite{TNSSMo}, and it is shown in Table~\ref{bps}. 
 Table~\ref{hdcy2} presents the branching ratios for the decay of $h_{125}$ for the three benchmark points that we have selected. Notice 
 that the ratios into the pseudoscalar pair $\mathcal{B}(h_{125}\to a_1 a_1)$ is about 10-20\%. The $a_1$ pair
 then decays into $b$ and $\tau$ pairs with rates shown in Table~\ref{a1dcy2}. We have selected a final state with $2b+2\tau$, where one of the $a_1$ decays into a $\tau$ pair  
and the other one decays into a $b$ pair. This also enhances the combinatorial factor and thus the number of events in the final state. The dominant SM backgrounds in this case comes from $t\bar{t}$, $ZZ$ and $b\bar{b}Z$.

 Figure~\ref{njtaupt} (right) shows that the requirement of a lower number of jets ($n_j$) $\leq 5$  will suppress the $t\bar{t}$ backgrounds. A similar effect is generated by requiring a lower $p_T$ on the $\tau_{\rm{jet}}$'s and $b_{\rm{jet}}$'s  ($p_T \lesssim 50$ GeV).  The corresponding $\tau$ decays give rise to very soft neutrinos, and therefore, by demanding a low missing $p_T$ $\leq 30$ GeV, we can reduce the backgrounds even further. The $b$ and $\tau$ tagging come with their own efficiencies \cite{btag} and \cite{taujet}, but this also helps in suppressing the other multi-jet backgrounds present from the SM.

In Table~\ref{2b2tau13} and Table~\ref{2b2tau14} we present the number of events for the three benchmark points coming both from the signal  and the SM backgrounds at the LHC, for a center of mass energy of 13 TeV and 14 TeV respectively.  The tables also show how their values change with each additional cut. We ask for a final state with $n_j\leq 5$, in which we demand the presence of at least two $b_{\rm{jet}}$'s and two $\tau_{\rm{jet}}$'s. In our notations, this request is indicated in the form: $n_j\leq 5\,[2b_{\rm{jet}}\,+ \,2\tau_{\rm{jet}}]$. We will be using the ampersand \& (a logical {\em and}) to combine additional constraints on the event, either in the form of particle/jet multiplicites or kinematical restrictions, and  define the signal as 
$$\rm{sig1}: n_j\leq 5\,[2b_{\rm{jet}}\, + \,2\tau_{\rm{jet}}]\,\&\,\ptmiss \leq 30 \, \rm{GeV.}$$
In the expression above, we have also required that the missing transverse momentum is smaller than 30 GeV 
$(\&\,\ptmiss \leq 30 \, \rm{GeV})$.
 In addition we apply some other cuts on the signal in order to reduce the backgrounds. For instance, in 
Table \ref{2b2tau13} we introduce a long sequence of such cuts (first column). In the case of BP1, for instance, the significance, after these selections, is $4.00\, \sigma$. The two additional conditions $p_1$ and $p_2$ are then applied as alternative clauses, and are enclosed into separate rows. \\
The first sequential cuts include  the $b_{\rm{jet}}$ pair invariant mass veto around $m_Z$,  the conditon that $|m_{bb}-m_Z|>10$ GeV and, around $m_{125}$, the condition$|m_{bb}-m_{h_{125}}|>10$ GeV.  $m_Z$ is the mass of the $Z$ gauge boson and $m_{h_{125}}$ is the Higgs mass (125 GeV).
Similarly, we also put veto on the invariant mass of the $\tau_{\rm{jet}}$ pair as: $|m_{\tau\tau}-m_Z|>10$ GeV and $|m_{\tau\tau}-m_{h_{125}}|>10$ GeV. Finally, since we are searching for hidden Higgs bosons, we demand that $m_{\tau\tau}<125$ GeV and $m_{bb}<125$ GeV respectively, where $m_{bb}$ and $m_{\tau\tau}$ are the invariant masses of the $b$ and $\tau$ pairs.

\begin{table}[t]
\begin{center}
\hspace*{-1.0cm}
\renewcommand{\arraystretch}{1.0}
\begin{tabular}{|c||c|c|c||c|c|c|c|c||}
\hline\hline
Final states&\multicolumn{3}{|c||}{Benchmark}&\multicolumn{5}{|c||}{Backgrounds }
\\
\hline
&BP1 & BP2&BP3 & $t\bar{t}$& $ZZ$ & $Z h$  &$b\bar{b}h$& $b\bar{b}Z$\\
\hline
\hline
$n_j\leq 5\,[2b_{\rm{jet}}+ 2\tau_{\rm{jet}}$]&\multirow{2}{*}{220.10}&\multirow{2}{*}{591.46}&\multirow{2}{*}{310.19}&\multirow{2}{*}{1824.08}&\multirow{2}{*}{199.50}&\multirow{2}{*}{39.56}&\multirow{2}{*}{11.87}&\multirow{2}{*}{4903.05}\\
$\&\,\ptmiss \leq 30$ GeV&&&&&&&&\\
&&&&&&&&\\
$\&\, p_T^{bj_{1,2}}\leq 50$GeV&\multirow{2}{*}{211.30}&\multirow{2}{*}{568.14}&\multirow{2}{*}{289.02}&\multirow{2}{*}{410.83}&\multirow{2}{*}{73.04}&\multirow{2}{*}{7.87}&\multirow{2}{*}{3.96}&\multirow{2}{*}{2941.83}\\
$\& \, |m_{bb}-m_Z|>10$ GeV&&&&&&&&\\
&&&&&&&&\\
$ \&\, |m_{bb}-m_{h_{125}}|>10$ GeV&211.30&565.32&289.02&386.18&73.04&7.52&3.96&2614.96\\
&&&&&&&&\\
$ \&\, |m_{\tau\tau}-m_Z|>10$ GeV&211.30&560.37&289.02&312.23&62.13&6.29&3.46&2397.04\\
&&&&&&&&\\
$\&\, |m_{\tau\tau}-m_{h_{125}}|>10$ GeV&211.30&560.37&289.02&287.58&62.13&6.18&2.97&2397.04\\
&&&&&&&&\\
$\&\, m_{\tau\tau}<125$GeV&211.30&560.37&289.02&254.71&62.13&6.18&2.97&2397.04\\
&&&&&&&&\\
$\&\, m_{bb}<125$GeV&211.30&559.66&289.02&230.06&62.13&6.07&2.97&2288.09\\
&&&&&&&&\\
\hline
Significance&4.00&9.98&5.39&\multicolumn{5}{|c||}{}\\
\hline
\hline
\multirow{3}{*}{\&\, $p_1:|m_{bb}-m_{a_1}|\leq 10$GeV}&\multirow{3}{*}{198.82}&\multirow{3}{*}{281.95}&\multirow{3}{*}{216.04}&24.65&0.00&0.22&0.49&326.87\\
&&&&65.73&26.16&1.46&0.49&1307.48\\
&&&&65.73&8.72&1.34&1.00&435.83\\
\hline
Significance&8.47&6.87&8.01&\multicolumn{5}{|c||}{}\\
\hline
\hline
%&&&&&&&&&\\
\multirow{3}{*}{$\&\, p_2:|m_{\tau\tau}-m_{a_1}|\leq 10$GeV}&\multirow{3}{*}{205.29}&\multirow{3}{*}{229.66}&\multirow{3}{*}{203.63}&65.73&3.27&0.33&0.00&0.00\\
&&&&73.95&28.34&1.46&0.49&762.70\\
&&&&41.08&13.08&1.57&1.48&0.00\\
\hline
Significance&12.40&6.94&12.65&\multicolumn{5}{|c||}{}\\
\hline
\hline
%&&&&&&&&&\\
\end{tabular}
\caption{The number of events for a $n_j\leq 5\,[2b_{{\rm{jet}}}+ 2\tau_{\rm{jet}}]\,\&\ptmiss \leq 30$ GeV final state at 100 fb$^{-1}$ of luminosity at the LHC, for a center of mass energy of 13 TeV. We require that the original signal has a number of jets 
$\leq 5$, of which 2 are $b_{\rm{jet}}$'s and 2 are $\tau_{\rm jet}$'s, with a missing $p_T\, (\ptmiss) \leq$ 30 GeV. We have denoted with $p_T^{bj_{1,2}}$ the transverse momentum of the $b_{\rm{jet}}$'s, with the two $b$'s labelled as 1 and 2. The final states are selected by imposing a long list of sequential cuts on the event, indicated with an ampersand (\&). The two additional options $p_1$ and $p_2$ are, however, alternative, and are imposed as additional constraints (a logical {\em or}). For this reason they are enclosed into separate rows.}\label{2b2tau13}
\end{center}
\end{table}
%%%%%%%%%%%%%%%%%%%%%%%%%%%%%%%%%%%%%%%%%%%%%%%%%%%%%%%

From Table~\ref{2b2tau13} and Table~\ref{2b2tau14} we deduce that the most dominant SM backgrounds are those from $t\bar{t}$, $ZZ$, $Zh$, $b\bar{b}h$ and $b\bar{b}Z$ respectively.
Though the 125 GeV bound on the two invariant masses reduces substantially most of the backgrounds, still the $b\bar{b}Z$ rate remains relatively large.  At this stage the signal significances, for the two benchmark points BP2 and BP3, both cross the $5\,\sigma$ value  
at an integrated luminosity 100 fb$^{-1}$, $9.98\, \sigma$ and $5.39 \,\sigma$, for a center of mass energy of 13 TeV.  In the case of BP1 this value is at the level of $4 \,\sigma$.  This is expected, given that in the case of BP2 the branching ratio $\mathcal{B}(h_{125}\to a_1a_1)$ is about $16\%$ (see Table~\ref{hdcy2}) and the 
pseudoscalar is relatively heavy, with a mass around $57$ GeV.  The  $\tau_{\rm jet}$'s and $b_{\rm{jet}}$'s coming from the decays of the $a_1$ are relatively harder (characterized by a larger momentum) compared to the benchmark points BP1 and BP3, so less events are cut out by the threshold on the $p_T$ cuts. Thus for BP2 we can reach a $5\sigma$ level of signal significance at an integrated luminosity of 25 fb$^{-1}$, for a given center of mass energy of 13 TeV. In this case the signal significance stays very similar also at 14 TeV, with little improvement for each of the $BP_i$'s. The signal significances, in this case, are  $4.47 \,\sigma$, $10.18\, \sigma$ and $5.98 \,\sigma$ respectively for BP1, BP2 and BP3.

%%%%%%%%%%%%%%%%%%%%%%%% 2b +2 \tau at 14 TeV%%%%%%%%%%%%%%%%%%%%
\begin{table}
\begin{center}
\hspace*{-1.0cm}
\renewcommand{\arraystretch}{1.0}
\begin{tabular}{|c||c|c|c||c|c|c|c|c||}
\hline\hline
Final states&\multicolumn{3}{|c||}{Benchmark}&\multicolumn{5}{|c||}{Backgrounds }\\
\hline
&BP1 & BP2&BP3 & $t\bar{t}$& $ZZ$ & $Z h$  &$b\bar{b}h$& $b\bar{b}Z$\\
\hline
\hline
$n_j\leq 5\,[2b_{\rm{jet}}+ 2\tau_{\rm{jet}}$]&\multirow{2}{*}{253.10}&\multirow{2}{*}{641.50}&\multirow{2}{*}{361.69}&\multirow{2}{*}{1530.66}&\multirow{2}{*}{223.72}&\multirow{2}{*}{40.35}&\multirow{2}{*}{19.77}&\multirow{2}{*}{4657.83}\\
$\&\,\ptmiss \leq 30$ GeV&&&&&&&&\\
&&&&&&&&\\
$p_T^{bj_{1,2}}\leq 50$ GeV&\multirow{2}{*}{248.41}&\multirow{2}{*}{605.68}&\multirow{2}{*}{337.04}&\multirow{2}{*}{294.36}&\multirow{2}{*}{85.11}&\multirow{2}{*}{7.80}&\multirow{2}{*}{7.19}&\multirow{2}{*}{3432.09}\\
$\&\, |m_{bb}-m_Z|>10$ GeV&&&&&&&&\\
&&&&&&&&\\
$\&\, |m_{bb}-m_{h_{125}}|>10$ GeV&248.41&604.89&337.04&294.36&85.11&7.43&7.19&3432.09\\
&&&&&&&&\\
$\&\, |m_{\tau\tau}-m_Z|>10$ GeV&248.41&597.73&337.04&255.11&70.52&6.09&5.39&2819.21\\
&&&&&&&&\\
$\&\, |m_{\tau\tau}-m_{h_{125}}|>10$ GeV&248.41&597.73&337.04&255.11&70.52&5.97&2.40&2819.21\\
&&&&&&&&\\
$\&\, m_{\tau\tau}<125$ GeV&248.41&596.93&337.04&255.11&69.30&5.85&2.40&2819.21\\
&&&&&&&&\\
$\&\, m_{bb}<125$ GeV&248.41&596.93&337.04&196.24&69.30&5.85&2.40&2574.07\\
&&&&&&&&\\
\hline
Significance&4.47&10.18&5.98&\multicolumn{5}{|c||}{}\\
\hline
\hline
\multirow{3}{*}{$\&\,p_1:|m_{bb}-m_{a_1}|\leq 10$ GeV}&\multirow{3}{*}{236.43}&\multirow{3}{*}{326.32}&\multirow{3}{*}{279.49}&9.81&2.43&0.37&0.00&490.30\\
&&&&68.68&31.61&1.83&1.20&1348.32\\
&&&&29.43&15.81&1.46&0.00&490.30\\
\hline
Significance&8.70&7.74&9.79&\multicolumn{5}{|c||}{}\\
\hline
\hline
\multirow{3}{*}{\&\,$p_2:|m_{\tau\tau}-m_{a_1}|\leq 10$ GeV}&\multirow{3}{*}{241.64}&\multirow{3}{*}{248.32}&\multirow{3}{*}{279.49}&19.62&6.08&0.49&0.00&0.00\\
&&&&58.87&24.32&1.58&0.00&1103.17\\
&&&&49.06&14.59&1.10&1.80&122.57\\
\hline
Significance&14.78&6.56&12.93&\multicolumn{5}{|c||}{}\\
\hline
\hline
\end{tabular}
\caption{The number of events for a $n_j\leq 5\,[2b_{{\rm{jet}}}+ 2\tau_{\rm{jet}}]\, \& \ptmiss \leq 30$ GeV final state at 100 fb$^{-1}$ of luminosity at the LHC for center of mass energy of 14 TeV. }\label{2b2tau14}.
\end{center}
\end{table}
%%%%%%%%%%%%%%%%%%%%%%%%%%%%%%%%%%
\begin{figure}[bht]
\begin{center}
\includegraphics[width=0.33\linewidth, angle=-90]{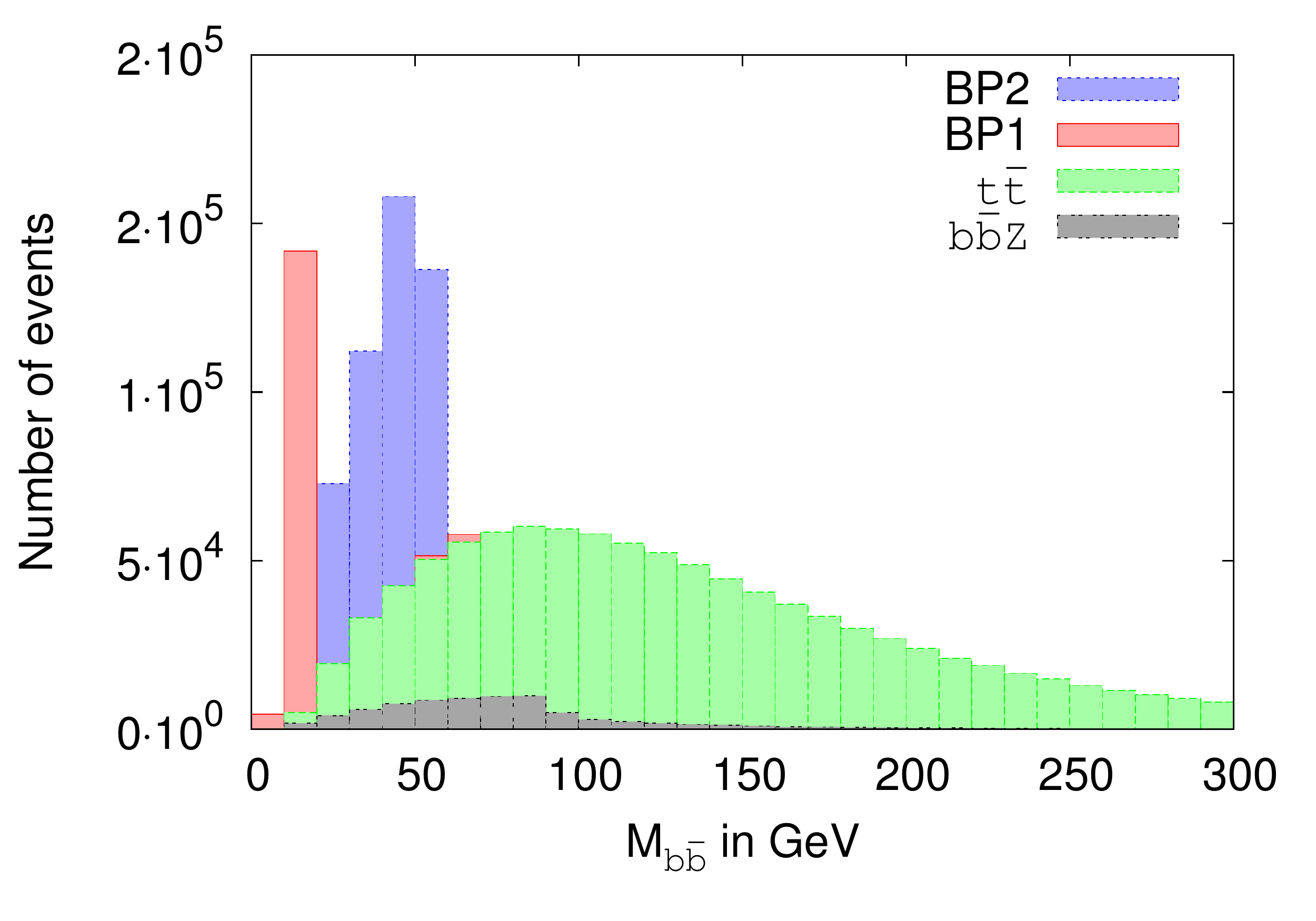}
\includegraphics[width=0.33\linewidth, angle=-90]{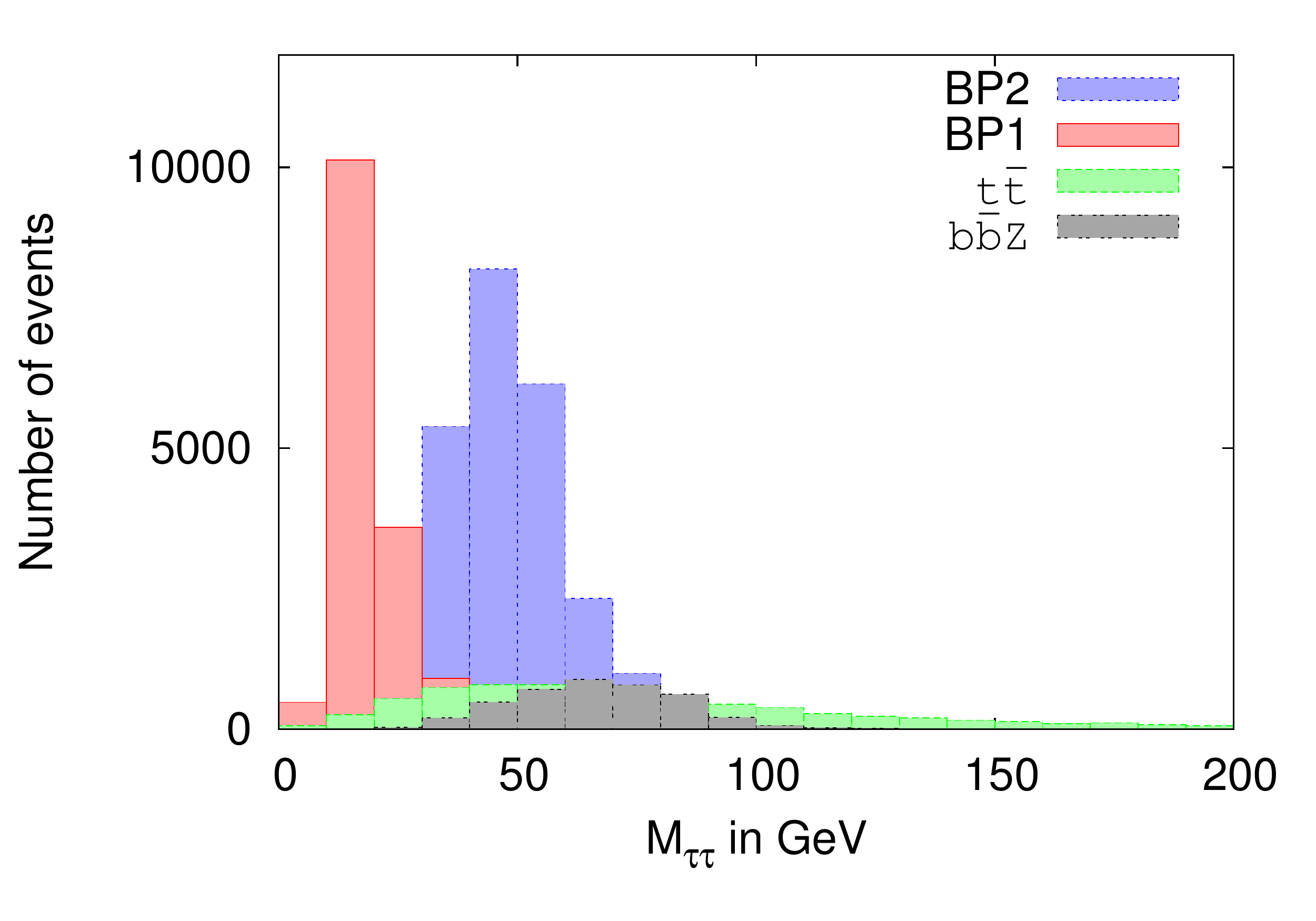}
\caption{ Invariant mass distribution of $b_{\rm{jet}}$'s (left) and $\tau_{\rm{jet}}$'s (right) for $t\bar{t}$ and for the signal in BP2.}\label{invdis}. 
\end{center}
\end{figure}
%%%%%%%%%%%%%%%%%%%%%%%%%%%%%%%%%%%%%%%%%%%%%%%%%%%%%%%%%%%%%%
Next we have analyzed the invariant mass distributions of the $b_{\rm{jet}}$ pair for the same benchmark points. Figure~\ref{invdis} (left) presents the $b_{\rm{jet}}$ pair invariant mass
distributions for the signal in BP1 and BP2, with dominant SM backgrounds coming from $t\bar{t}$ and $b\bar{b}Z$. These results  suggest that, given
the integrated luminosity, it is possible to resolve the resonant peak in the mass distribution of the signal. To further clarify this point, we select events with $|m_{bb}-m_{a_1}|\leq 10$ GeV, that we label as $p_1$.
The resolutions of these peaks depend on the specific benchmark point, but this selection reduces the $b\bar{b}Z$ background drastically, in those cases when $m_{a_1}$  is far separated from the $Z$ gauge boson mass $m_Z$.
The signal significances for all the benchmark points cross the $5\sigma$ level at an integrated luminosity of 100 fb$^{-1}$, and at 13 TeV they are equal to $8.47\, \sigma$,  $6.87\, \sigma$ and $8.01\, \sigma$  for 
BP1, BP2 and BP3 respectively. At a center of mass energy of 14 TeV the significances are $8.70\, \sigma$, $7.74 \,\sigma$ and $9.79 \,\sigma$ in the three cases.

Finally, we simulate the $\tau_{\rm{jet}}$ invariant mass distributions, as they are expected to be cleaner than the $b_{\rm{jet}}$ distributions. Figure~\ref{invdis} (right) shows the invariant mass distributions for both the signals in BP1 and BP3, and the SM backgrounds from $t\bar{t}$ and $b\bar{b}Z$. For this purpose, similarly to the previous case, we select those events with $|m_{\tau\tau}-m_{a_1}|\leq 10$ GeV. For the points which are far away from the $Z$ mass, namely BP1 and BP3, the
signal significance improves significantly, to $12.40 \,\sigma$ and $12.65 \,\sigma$ respectively, whereas for BP2 it is $6.94 \,\sigma$. At a centre of mass energy of 14 TeV these value
 are $14.78\, \sigma$, $6.56 \,\sigma$ and $12.93 \,\sigma$ for BP1, BP2 and BP3 respectively.

\subsection{$3\tau$}
In this subsection we consider the case in which both pseudoscalars decay into $\tau$ pairs.
In this case we expect to see a final state of $4\tau$' s. Of course, due to the lower branching ratio
 in the $a_1\to \tau\bar{\tau}$ mode, the final state numbers are not very promising at low luminosities.
 On top of that, due to a low $\tau$-tagging efficiency for $\tau$'s of low $p_T$, { the final state number is  furtherly reduced.\cite{taujet}.
 Keeping this in mind, we search for final states where we have at least three $\tau$'s. We tag such $\tau$'s
via hadronic $\tau_{\rm{jet}}$'s, as explained earlier.
%%%%%%%%%%%%%%%%% >=3 \tau at 13 TeV  %%%%%%%%%%%%%%%%%%%%
\begin{table}[h]

\begin{center}
\hspace*{-1.0cm}
\renewcommand{\arraystretch}{1.2}
\begin{tabular}{|c||c|c|c||c|c|c||}
\hline\hline
Final states&\multicolumn{3}{|c||}{Benchmark}&\multicolumn{3}{|c||}{Backgrounds }
\\
\hline
&BP1 & BP2&BP3  &$ZZ$  &$ZW^\pm$&$h Z$\\
\hline
\hline
$n_j\leq 5\,[\geq 3\tau_{\rm{jet}}]$&95.71&199.27&137.21&186.42&437.17&20.68\\
&&&&&&\\
$\&\, |m_{\tau\tau}-m_Z|>10$ GeV&94.79&197.15&135.02&163.53&363.43&17.42\\
&&&&&&\\
$\&\, m_{\tau\tau}\leq125$ GeV&94.79&197.15&135.02&158.07&326.56&16.07\\
&&&&&&\\
 \&\, $p_T^{\tau_{j_1}}\leq 100\, \&\, p_T^{\tau_{j_{2,3}}}\leq 50$ GeV&87.85&184.43&123.34&99.21&210.69&8.31\\
\hline
Significance&4.41&8.22&5.93&\multicolumn{3}{|c||}{}\\
\hline\hline
\multirow{3}{*}{$\&\, p_1:|m_{\tau\tau}-m_{a_1}|\leq 10$ GeV}&\multirow{3}{*}{48.55}&\multirow{3}{*}{54.41}&\multirow{3}{*}{64.96}&4.36&21.07&0.90\\
&&&&44.70&89.54&2.70\\
&&&&26.16&42.14&3.82\\
\hline
Significance&5.61&3.93&5.55&\multicolumn{3}{|c||}{}\\
\hline
\hline
\end{tabular}
\caption{The number of events for a $n_j\leq 5\,[\geq 3\tau_{\rm{jet}}]$ final state at 100 fb$^{-1}$ of luminosity at the LHC with 13 TeV center of mass energy.}\label{3tau13}
\end{center}
\end{table}
%%%%%%%%%%%%%%%%%%%%%%%%%%%%%%%%%%%%%%%%%%%%%%%%%%%%%%%%
The dominant SM backgrounds, in this case, come from the association of $Z$ bosons, i.e. from $ZZ$, $ZW^\pm$, $Zh$
  along with the triple gauge boson productions, namely from $ZZZ$, $ZZW^\pm$, $W^\pm W^\mp W^\pm$, $ZW^\pm W^\mp$ and $WWW$.
  However, the triple gauge boson backgrounds are found to be negligible after imposing the cuts ($\lsim 0.1$) at 100 fb$^{-1}$.
 Table~\ref{3tau13} and Table~\ref{3tau14} show the expected numbers of events for the three benchmark points $BP_i$, together with the dominant backgrounds, at an integrated luminosity of 100 fb$^{-1}$. The final state that we are looking for is characterized by a number of jets $n_j\leq 5$ among which we tag at least  three of them as $\tau_{\rm{jet}}$'s, defined as
 $$\textrm{sig}2: n_j\leq 5\,[\geq 3\tau_{\rm{jet}}].$$
 
 We then add some further kinematical
 cuts to reduce the backgrounds, as before. These cuts include the invariant mass veto on the $\tau_{\rm{jet}}$ pair, $|m_{\tau\tau}-m_Z|>10$ GeV and we also demand that $m_{\tau\tau}\leq125$ GeV, which allows us to search for hidden resonances. Finally, we also demand for softer second and third $\tau_{\rm{jet}}$'s by implementing the cuts $p_T^{\tau_{j_1}}\leq 100\, \&\, p_T^{\tau_{j_{2,3}}}\leq 50$ GeV.

From Table~\ref{3tau13} and Table~\ref{3tau14} one deduces that the $ZW^\pm$ channel remains the most dominant background of all. The signal significance 
at this stage for the three benchmark points are $4.41 \,\sigma$, $8.22\, \sigma$ and $5.93\, \sigma$ for BP1, BP2 and BP3 respectively, at an integrated luminosity of 100 fb$^{-1}$ and a center of mass energy of 13 TeV. At 14 TeV these numbers are  $3.79 \,\sigma$, $8.38 \, \sigma$ and $5.81\, \sigma$.

As in the previous case, also in this case we try to select events around the pseudoscalar mass peak by the constraint $p_1:|m_{\tau\tau}-m_{a_1}|\leq 10$ GeV. The mass resolution depends on the mass value of $a_1$, but BP1 and BP3 now have more than a
$5\sigma$ signal significance. For BP2 $m_{a_1} \sim 57$ GeV, and the multiplicities from the backgrounds involving $ZZ$ and $ZW^\pm$
are more significant than for BP1 and BP3.
 The signal significance at 13 TeV, with an integrated luminosity of 100 fb$^{-1}$ for BP1, BP2 and BP3 are $5.61\, \sigma$, $3.93\, \sigma$ and $5.55\, \sigma$
respectively. These values change for collisions at 14 TeV and equal $5.16\, \sigma$, $4.00 \,\sigma$ and $6.03\, \sigma$ in this second case.
%%%%%%%%%%%%%%%%%%%%%% 14 TeV %%%%%%%%%%%%%%%%%%%%%%%%%%%%%%
\begin{table}

\begin{center}
\hspace*{-1.0cm}
\renewcommand{\arraystretch}{1.2}
\begin{tabular}{|c||c|c|c||c|c|c||}
\hline\hline
Final states&\multicolumn{3}{|c||}{Benchmark}&\multicolumn{3}{|c||}{Backgrounds }
\\
\hline
&BP1 & BP2&BP3 & $ZZ$  &$ZW^\pm$&$h Z$\\
\hline
\hline
$n_j\leq 5\,[\geq 3\tau_{\rm{jet}}]$&96.34&224.45&146.73&200.62&499.20&18.28\\
&&&&&&\\
$\&\, |m_{\tau\tau}-m_Z|>10$ GeV&94.78&222.85&142.62&178.73&408.70&15.11\\
&&&&&&\\
$\&\, m_{\tau\tau}\leq125$ GeV&94.78&222.06& 141.80&165.36&382.43&13.65\\
&&&&&&\\
$\&\, p_T^{\tau_{j_1}}\leq100\,\& \,p_T^{\tau_{j_{2,3}}}\leq 50$ GeV&82.80&205.34& 133.58&121.59&265.66&7.56\\
\hline
Significance&3.79&8.38&5.81&\multicolumn{3}{|c||}{}\\
\hline\hline
\multirow{3}{*}{$\&\, p_1:|m_{\tau\tau}-m_{a_1}|\leq 10$ GeV}&\multirow{3}{*}{46.35}&\multirow{3}{*}{62.08}&\multirow{3}{*}{79.74}&12.16&20.44&1.71\\
&&&&54.71&122.61&2.44\\
&&&&25.53&67.14&2.56\\
\hline
Significance&5.16&4.00&6.03&\multicolumn{3}{|c||}{}\\
\hline
\hline
\end{tabular}
\caption{The number of events for a $n_j\leq 5\,[\geq 3\tau_{\rm{jet}}]$ final state at 100 fb$^{-1}$ of luminosity at the LHC, for a center of mass energy of 14 TeV.  }\label{3tau14}
\end{center}
\end{table}
%%%%%%%%%%%%%%%%%%%%%%%%%%%%%%%%%%

\subsection{$2b+2\mu$}
The decay rate of the pseudoscalar  to $\mu\bar{\mu}$ is $\mathcal{O}(10^{-4})$, which
makes this channel difficult to observe. If we demand that one of the two pseudoscalars decay into a $b\bar{b}$
pair and the other into a $\mu\bar{\mu}$ pair, the effective cross-section may increase firstly due to
the large branching coming from $a_1\to b\bar{b}$ and, secondly, due to a combinatorial factor of 2, because of the presence of two pseudoscalars. This gives us the option of investigating a final state $2b+2\mu$. 

Table~\ref{2b2mu13} and Table~\ref{2b2mu14} show the corresponding $2\mu$ final states event numbers 
for the benchmark points and the dominant SM backgrounds which include $t\bar t$, $ZZ$, $Zh$, $b\bar b h$ and $b \bar b Z$
at an integrated luminosity of 1000 fb$^{-1}$. We first consider the $2\mu \,\&\, p_T^{\ell_{1,2}}\leq 50$ GeV final state, largely dominated by the SM backgrounds (see Tables~\ref{2b2mu13} and~\ref{2b2mu14}).  Then with impose further requirements on the numbers of jets and their transverse momentum ($p_T$), by defining the signal as 
 $$\textrm{sig}3: \, n_j\leq 3\,[2b_{\rm jet}]\, \& \,n_{\mu}\geq 2 \,[|m_{\mu\mu}-m_Z|> 5 \,\rm{GeV}]\, \& \,p_T^{{\mu,j}_{1,2}}\leq 50\, \rm{GeV}\, \& \,\ptmiss \leq 30\, \textrm{GeV}.$$ 
 
The $\mu$-pair invariant mass veto around the $Z$ mass $(|m_{\mu\mu}-m_Z|> 5 \,\rm{GeV})$, together with the condition of having softer $b_{\rm{jet}}$'s in the final state ($p_T^{j_{1,2}}\leq 50 \,\rm{GeV}$), conspire to reduce the SM backgrounds coming from the $Z$ bosons quite drastically. Finally, since this final state - in an ideal situation - should not have any missing energy, we also demand that
$\ptmiss \leq 30$ GeV. To reduce the backgrounds even further, and to ensure that we select signatures of the light pseudoscalar decay below $125$ GeV, we impose additional constraints on the $\mu$-pair and on the $b_{\rm{jet}}$-pair invariant masses, around the $Z$ mass and the mass of $h_{125}$. These are given by
$|m_{\mu\mu}-m_{h_{125}}|>5\, \textrm{ GeV}$, $|m_{bb}-M_Z|\geq 10\, \textrm{ GeV}$ and |$m_{bb}-m_{h_{125}}|>10 \, \textrm{ GeV}$.\\
At this stage, only in the case of BP2 the signal significance reaches the $3.31 \,\sigma$ value, while for BP1 and BP3 these are $1.03 \,\sigma$,  and $1.83 \,\sigma$ respectively, at 13 TeV. At a center of mass energy of 14 TeV, instead, the values are $1.08 \,\sigma$, $2.64\, \sigma$ and $1.18\, \sigma$ respectively for BP1, BP2 and BP3. 
%%%%%%%%%%%%%%%%% 2 b+2\mu at 13 TeV  %%%%%%%%%%%%%%%%%%%%
\begin{table}[t]
\begin{center}
\hspace*{-2cm}
\renewcommand{\arraystretch}{1.4}
\begin{tabular}{|c||c|c|c||c|c|c|c|c||}
\hline\hline
Final states&\multicolumn{3}{|c||}{Benchmark}&\multicolumn{5}{|c||}{Backgroounds }\\
\hline
&BP1 & BP2&BP3 & $t\bar t$&$ZZ$& $Zh$ & $b\bar b h$  &$b \bar b Z$\\
\hline
\hline
$2\mu_{\rm{jet}}\, \& \, p_T^{\ell_{1,2}}\leq 50$ GeV& 1877.23&3660.42&3167.55&909080&132161&2669.20&657.71&$6.3\times10^6$\\
&&&&&&&&\\
$\&\,n_j\leq 3\,\&\, b_{\rm jet}\geq 2$&\multirow{3}{*}{69.36}&\multirow{3}{*}{226.13}&\multirow{3}{*}{124.07}&\multirow{3}{*}{4765.60}&\multirow{3}{*}{457.87}&\multirow{3}{*}{15.73}&\multirow{3}{*}{14.83}&\multirow{3}{*}{28.60}\\
$\&\,|m_{\mu\mu}-m_Z|> 5$ GeV&&&&&&&&\\
$\&\,p_T^{j_{1,2}}\leq 50 \rm{GeV}\,\&\, \ptmiss \leq 30$ GeV&&&&&&&&\\
$\&\, |m_{\mu\mu}-m_{h_{125}}|>5$ GeV&\multirow{2}{*}{69.36}&\multirow{2}{*}{226.13}&\multirow{2}{*}{124.07}&\multirow{2}{*}{4190.45}&\multirow{2}{*}{359.76}&\multirow{2}{*}{14.61}&\multirow{2}{*}{14.83}&\multirow{2}{*}{28.60}\\
$\&\,|m_{bb}-M_Z|\geq 10$ GeV&&&&&&&&\\
$\&\, |m_{bb}-m_{h_{125}}|>10$ GeV&69.36&226.13&124.07&4026.11&359.76&13.49&14.83&28.60\\
\hline
Significance&1.03&3.31&1.83&\multicolumn{5}{|c||}{}\\
\hline
\hline
\multirow{3}{*}{$\&\, p_1:|m_{bb}-m_{a_1}|\leq 10$ GeV}&\multirow{3}{*}{64.73}&\multirow{3}{*}{98.93}&\multirow{3}{*}{80.28}&328.66&0.00&0.00&4.94&19.67\\
&&&&1150.32&141.72&5.62&9.89&9.53\\
&&&&492.99&43.61&2.25&0.00&0.00\\
\hline
Significance&3.17&2.63&3.23&\multicolumn{5}{|c||}{}\\
\hline
\hline
\multirow{3}{*}{$\&\, p_2:|m_{\mu\mu}-m_{a_1}|\leq 5$ GeV}&\multirow{3}{*}{41.61}&\multirow{3}{*}{148.40}&\multirow{3}{*}{72.98}&328.66&43.61&1.12&0.00&0.00\\
&&&&575.15&32.70&0.00&0.00&9.53\\
&&&&410.83&21.80&1.12&4.94&0.00\\
\hline
Significance&2.04&5.36&3.22&\multicolumn{5}{|c||}{}\\
\hline
\hline
\end{tabular}
\caption{The number of events for the $n_j\leq 3\,[2b_{\rm{jet}}]\, \&\, \geq 2\mu\, \& \,\ptmiss \leq 30$ GeV final state at 1000 fb$^{-1}$ of luminosity at the LHC, for a center of mass energy of 13 TeV. The constraint
$(\& \geq 2\mu)$ requires the presence of at least 2 muons. The clause ($\&\, b_{\rm jet}\geq 2$) demands at least 2  jets of $b$ quarks, denoted as $b_{\rm jet}$. }\label{2b2mu13}
\end{center}
\end{table}
%%%%%%%%%%%%%%%%%%%%%%%%%%%%%%%%%%
Later we try to enhance the mass peak resolutions on the $bb$ and $\mu\mu$ invariant mass distributions by imposing the two 
constraints (denotes as $p_1,p_2$)
$$p_1:|m_{bb}-m_{a_1}|\leq 10\, \textrm{GeV \,\, and}\,\, p_2:|m_{\mu\mu}-m_{a_1}|\leq 5 \,\textrm{GeV}.$$ At a center of mass energy of 13 TeV, the $m_{bb}$ peaks are characterized by about a $3\, \sigma$ signal significance i.e., $3.17\sigma$, $2.63 \, \sigma$ and $3.23\, \sigma$ respectively for BP1, BP2 and BP3 at an integrated luminosity of of 1000 fb$^{-1}$. At 14 TeV the respective values are $3.17\, \sigma$, $2.63 \, \sigma$ and $3.23 \, \sigma$ respectively for the three benchmarks. \\
The constraint $p_2:|m_{\mu\mu}-m_{a_1}|\leq 5$ GeV, brings BP2 at $5.36 \,\sigma$, BP1 at $2.04\, \sigma$, and BP3 at $3.22 \,\sigma$, for a center of mass energy of 13 TeV. At 14 TeV
the significances are $4.71\, \sigma$, $3.82 \, \sigma$ and $3.00 \, \sigma$ in the three cases, respectively. 
%%%%%%%%%%%%%%%%%% 2 b+2\mu at 14 TeV %%%%%%%%%%%%%%%%
\begin{table}
\begin{center}
\hspace*{-2cm}
\renewcommand{\arraystretch}{1.4}
\begin{tabular}{|c||c|c|c||c|c|c|c|c||}
\hline\hline
Final states&\multicolumn{3}{|c||}{Benchmark}&\multicolumn{5}{|c||}{Backgrounds }
\\
\hline
&BP1 & BP2&BP3 & $t\bar t$&$ZZ$& $Zh$ & $b\bar b h$  &$b\bar b Z$\\
\hline
\hline
$2\mu_{\rm{jet}}\,\&\, p_T^{\ell_{1,2}}\leq 50$ GeV& \multirow{2}{*}{2281.00}&\multirow{2}{*}{4011.37}&\multirow{2}{*}{3362.13}&\multirow{2}{*}{788683}&\multirow{2}{*}{141428}&\multirow{2}{*}{2926.71}&\multirow{2}{*}{946.42}&\multirow{2}{*}{$7\times10^6$}\\
&&&&&&&&\\
$\&\,n_j\leq 3\,\&\, b_{\rm jet}\geq 2$&\multirow{3}{*}{67.70}&\multirow{3}{*}{167.14}&\multirow{3}{*}{ 73.99}&\multirow{3}{*}{5102.21}&\multirow{3}{*}{583.61}&\multirow{3}{*}{20.72}&\multirow{3}{*}{17.97}&\multirow{3}{*}{10.72}\\
$\&\,|m_{\mu\mu}-m_Z|> 5$ GeV&&&&&&&&\\
$\&\,p_T^{j_{1,2}}\leq 50 \rm{GeV}\,\&\, \ptmiss \leq 30$ GeV&&&&&&&&\\
$|m_{\mu\mu}-m_{h_{125}}|>5$ GeV&\multirow{2}{*}{67.70}&\multirow{2}{*}{167.14}&\multirow{2}{*}{73.99}&\multirow{2}{*}{3630.42}&\multirow{2}{*}{510.66}&\multirow{2}{*}{9.75}&\multirow{2}{*}{11.98}&\multirow{2}{*}{0.00}\\
$\&\,|m_{bb}-M_Z|\geq 10$ GeV&&&&&&&&\\
$\&\,|m_{bb}-m_{h_{125}}|>10$ GeV&67.70&167.14&73.99&3336.06&498.50&9.75&11.98&0.00\\
\hline
Significance&1.08&2.64&1.18&\multicolumn{5}{|c||}{}\\
\hline
\hline
\multirow{3}{*}{$\&\, p_1:|m_{bb}-m_{a_1}|\leq 10$ GeV}&\multirow{3}{*}{67.70}&\multirow{3}{*}{79.60}&\multirow{3}{*}{ 57.54}&196.24&0.00&0.00&0.00&0.00\\
&&&&1373.67&255.33&1.22&0.00&0.00\\
&&&&686.83&24.32&2.44&0.00&0.00\\
\hline
Significance&4.16&1.93&2.08&\multicolumn{5}{|c||}{}\\
\hline
\hline
\multirow{3}{*}{$\&\, p_2:|m_{\mu\mu}-m_{a_1}|\leq 5$ GeV}&\multirow{3}{*}{41.66}&\multirow{3}{*}{103.47}&\multirow{3}{*}{ 45.21}&0.00&36.47&0.00&0.00&0.00\\
&&&&588.72&36.47&0.00&5.99&0.00\\
&&&&98.12&85.11&0.00&0.00&0.00\\
\hline
Significance&4.71&3.82&3.00&\multicolumn{5}{|c||}{}\\
\hline
\hline
\end{tabular}
\caption{The number of events for $n_j\leq 3\,[2b_{\rm{jet}}]\, \&\, \geq 2\mu\, \&\,\ptmiss \leq 30$ GeV final state at 1000 fb$^{-1}$ of luminosity at the LHC for center of a center of mass energy of 14 TeV.}\label{2b2mu14}
\end{center}
\end{table}
%%%%%%%%%%%%%%%%%%%%%%%%%%%%%%%%%%

\subsection{$2\tau+2\mu$}
In this section we discuss a scenario where one of the pseudoscalars decays into a $\tau$ pair and the second one into a $\mu$ pair. Due to the low branching ratios of these two modes, even with a large integrated luminosity, the signal remains small. It is however accompanied by a SM backgrounds for such final states ($2\tau+2\mu$) which is quite suppressed. As in the previous cases, also in this case we tag the $\tau$ via its hadronic decay into a $\tau_{\rm{jet}}$ \cite{taujet}.
The  threshold $p_T$ cuts both for the $\tau_{\rm{jet}}$ and for the muons are kept as low as 10 GeV, since we are considering the decay of a very light pseudoscalar.

%%%%%%%%%%%%%%%%% 2 \tau +2\mu at 13 TeV %%%%%%%%%%%%%%%%%%%%
\begin{table}
\begin{center}
\hspace*{-1cm}
\renewcommand{\arraystretch}{1.4}
\begin{tabular}{|c||c|c|c||c|c||}
\hline\hline
Final states&\multicolumn{3}{|c||}{Benchmark}&\multicolumn{2}{|c||}{Backgrounds }
\\
\hline
&BP1 & BP2&BP3 & $ZZ$& $Zh$ \\
\hline
\hline
$2\mu\, \&\, n_j\leq 3\,[2\tau_{\rm{jet}}$]&\multirow{2}{*}{16.18}&\multirow{2}{*}{14.13}&\multirow{2}{*}{29.19}&\multirow{2}{*}{490.58}&\multirow{2}{*}{28.10}\\
$\&\,p_T^{\ell_{1,2}}\,\&\,p_T^{j_{1,2}}\leq 50$ GeV&&&&&\\
$\&\,|m_{\mu\mu}-m_Z|\geq 5$ GeV&16.18&14.13&29.19&218.03&9.00\\
$\&\,|m_{\tau\tau}-m_Z|>10$ GeV&16.18&14.13&29.19&163.53&9.00\\
$\&\,|m_{\tau\tau}|<125$ GeV&16.18&14.13&29.19&152.62&7.87\\
\hline
Significance&1.22&1.07&2.12&\multicolumn{2}{|c||}{}\\
\hline
\hline
\multirow{3}{*}{$\&\,p_1:|m_{\tau\tau}-m_{a_1}|\leq 10$ GeV}&\multirow{3}{*}{11.56}&\multirow{3}{*}{14.13}&\multirow{3}{*}{21.90}&0.00&0.00\\
&&&&54.51&1.12\\
&&&&32.70&1.12\\
\hline
Significance&3.40&1.70&2.93&\multicolumn{2}{|c||}{}\\
\hline
\hline
\multirow{3}{*}{$\&\, p_2:|m_{\mu\mu}-m_{a_1}|\leq 5$ GeV}&\multirow{3}{*}{6.94}&\multirow{3}{*}{7.07}&\multirow{3}{*}{0.00}&0.00&0.00\\
&&&&0.00&0.00\\
&&&&43.61&2.25\\
\hline
Significance&2.63&2.65&-&\multicolumn{2}{|c||}{}\\
\hline
\hline
\end{tabular}
\caption{The number of events for $n_j\leq 3\,[2\tau_{\rm{jet}}] \,\&\,\geq 2\mu \,\&\,\ptmiss \leq 30$ GeV final state at 1000 fb$^{-1}$ of luminosity at the LHC for a center of mass energy of 13 TeV.}\label{2ta2mu13}
\end{center}
\end{table}
%%%%%%%%%%%%%%%%%%%%%%%%%%%%%%%%%%%%%%%%%%%%%%%%%%
The results of this analysis are reported in Table~\ref{2ta2mu13} and Table~\ref{2ta2mu14}, where we present the number of events for the benchmark points and the dominant SM backgrounds, for a center of mass energy of 13 and 14 TeV and an integrated luminosity of 1000 fb$^{-1}$. We search for a muon pair and at least two $\tau$'s in the final state. Though muons ($\mu$) will be detected as a charged leptons, the $\tau$'s will be 
detected via their hadronic decays as $\tau_{\rm{jets}}$'s \cite{taujet}. Being the two pseduoscalars light, we require  both the $\mu$ and the $\tau$ jets to be rather soft (i.e. ($p_T^{\ell_{1,2}}\&p_T^{j_{1,2}}) \leq 50$ GeV) in the final state. This defines the signal as 
$$\textrm{sig}4: \,n_j\leq 3\,[2\tau_{\rm{jet}}]\, \& \, \geq 2\mu \, \&\, \ptmiss \leq 30\, \rm{GeV}.$$

 Tagging both muons and requiring the cut $p_T \leq 50$ GeV for the transverse momentum $p_T$ of the $\tau_{\rm{jet}}$, will suppress much of the hard SM backgrounds, favouring the search for a low mass resonance, in this case a light 
pseudoscalar. The dominant backgrounds in this case comes from the SM $ZZ$ and $hZ$ channels. The background due to the $a_1 Z$ channel is negligible, due to the mostly-singlet nature of the $a_1$. We have also checked for other triple gauge boson contributions to this final states, but they are all either zero or negligible.  To reduce further the SM backgrounds we apply a veto on the mass peak of the $Z$ boson, by requiring that $|m_{\mu\mu}-m_Z|\geq 5$ GeV and $|m_{\tau\tau}-m_Z|>10$ GeV respectively.  As one may deduce from Table~\ref{2ta2mu13} and Table~\ref{2ta2mu14},  the application of these two cuts, though reduces the SM backgrounds quite drastically, does not affect the signal, which remains unchanged. Finally, we apply the constraint $|m_{\tau\tau}|<125$ GeV to ensure the search for
hidden scalars, i.e., $m_{a_1}< 125$ GeV, which causes an even larger suppression of the background. At this level the signal significances are still below $3\sigma$ at 13 TeV and reach $3.20 \, \sigma$ only in the case of the benchmark point BP3, at 14 TeV. 

Next we apply the constraint $p_1:|m_{\tau\tau}-m_{a_1}|\leq 10$ GeV to favour the search for a possible mass peak of the pseudoscalar and this enhances the signal significance to $3.40\, \sigma$, $1.70\, \sigma$ and $2.93\, \sigma$ respectively for BP1, BP2 and BP3 at 13 TeV. At 14 TeV these numbers are $2.47 \,\sigma$, $2.51\, \sigma$ and $3.27\, \sigma$ respectively. Similar peaks around $\mu$ pair invariant mass distribution, i.e. with $p_2:|m_{\mu\mu}-m_{a_1}|\leq 5$ GeV, give signal significances of $2.63 \,\sigma$ and $2.65\, \sigma$ for BP1 and BP2, at a center of mass energy of 13 TeV. BP3 in this case runs out of statistics. At 14 TeV the signal significances are $2.05 \,\sigma$, $2.82\, \sigma$ and $2.04 \,\sigma$ respectively. The leptonic modes thus need higher luminosities $\gsim 2000$ fb$^{-1}$ in order to reach the discover limit for a light pseudoscalar.
%%%%%%%%%%%%%%%%% 2 \tau +2\mu at 14 TeV %%%%%%%%%%%%%%%%%%%%
\begin{table}
\begin{center}
\hspace*{-1cm}
\renewcommand{\arraystretch}{1.4}
\begin{tabular}{|c||c|c|c||c|c||}
\hline\hline
Final states&\multicolumn{3}{|c||}{Benchmark}&\multicolumn{2}{|c||}{Backgrounds }
\\
\hline
&BP1 & BP2&BP3 & $ZZ$& $Zh$\\
\hline
\hline
$2\mu\,\&\, n_j\leq 3\,[2\tau_{\rm{jet}}$]&\multirow{2}{*}{15.62}&\multirow{2}{*}{31.84}&\multirow{2}{*}{41.10}&\multirow{2}{*}{498.50}&\multirow{2}{*}{20.72}\\
$\&\,p_T^{\ell_{1,2}}\,\&\,p_T^{j_{1,2}}\leq 50$ GeV&&&&&\\
$\&\,|m_{\mu\mu}-m_Z|\geq 5$ GeV&15.62&31.84&41.10&145.90&7.31\\
$\&\, |m_{\tau\tau}-m_Z|>10$ GeV&15.62&31.84&41.10&121.58&3.66\\
$\&\, |m_{\tau\tau}|<125$ GeV&15.62&31.84&41.10&121.58&2.44\\
\hline
Significance&1.32&2.55&3.20&\multicolumn{2}{|c||}{}\\
\hline
\hline
\multirow{3}{*}{$\&\,p_1:|m_{\tau\tau}-m_{a_1}|\leq 10$ GeV}&\multirow{3}{*}{15.62}&\multirow{3}{*}{15.92}&\multirow{3}{*}{28.77}&24.32&0.00\\
&&&&24.32&0.00\\
&&&&48.63&0.00\\
\hline
Significance&2.47&2.51&3.27&\multicolumn{2}{|c||}{}\\
\hline
\hline
\multirow{3}{*}{$\&\,p_2:|m_{\mu\mu}-m_{a_1}|\leq 5$ GeV}&\multirow{3}{*}{5.21}&\multirow{3}{*}{7.96}&\multirow{3}{*}{12.33}&0.00&1.22\\
&&&&0.00&0.00\\
&&&&24.32&0.00\\
\hline
Significance&2.05&2.82&2.04&\multicolumn{2}{|c||}{}\\
\hline
\hline
\end{tabular}
\caption{The number of events for $n_j\leq 3\,[2\tau_{\rm{jet}}]\,\&\,\geq 2\mu\,\&\,\ptmiss \leq 30$ GeV final state at 1000 fb$^{-1}$ of luminosity at the LHC for center of mass energy (ECM) of 14 TeV.}\label{2ta2mu14}
\end{center}
\end{table}
%%%%%%%%%%%%%%%%%%%%%%%%%%%%%%%%%%

\section{Discussions and conclusions}\label{concl}
In this article we have analysed signatures of a supersymmetric extension of the SM, characterized by an extra $Y=0$ Higgs triplet and a SM gauge singlet, in view of the recent and previous Higgs data. In particular, we have investigated the discovery potential of a light pseudoscalar sector which is present in this model.  Our analysis has been performed assuming as a production mechanism the gluon-gluon fusion channel of the 125 GeV Higgs $h_{125}$, and focused on the currents experimental rates on its decay into the $WW^*$, $ZZ^*$
and $\gamma\gamma$ derived at the LHC. Given the current uncertainties in these discovered modes as well as in other (fermionic) modes of the Higgs, we have investigated the possibility that such uncertainties are compatible with the production of two light pseudoscalars, predicted by the TNMSSM, which have so far been undetected.

Benchmarking three points in the parameter space of the model, we have proposed and simulated final states of the form $2b+2\tau$, $3\tau$, $2b+2\mu$ and $2\tau +2\mu$, derived from the decays of such pseudoscalars.
A PYTHIA-FastJet based simulation of the dominant SM backgrounds shows that, depending on the benchmark points, such light pseudoscalars can be probed with 
early LHC data ($\sim 25$ fb$^{-1}$) at 13 and 14 TeV. The $2\tau+2\mu$ decay modes of such states, though much cleaner compared to other channels, need higher luminosity ($\sim 2000$ fb$^{-1}$) in order to be significant. Nevertheless, such muon final states will be crucial for precision mass measurements of the $a_1$. In this case, due to the $Z-a_1-a_1$ coupling, one may consider the production of an $a_1$ pair directly at tree-level, and this can enhance the signal strength by about $10\%$.

The identification of such hidden scalars would be certainly a signal in favour of an extended Higgs sectors, but finding the triplet and singlet $SU(2)$ representations of these extra states would require more detailed searches. Clearly, there are some other distinctive features of this model respect to the NMSSM. The NMSSM does not have any extra charged Higgs bosons compared to the MSSM, while the TNMSSM has an extra triplet-like charged Higgs boson which does not couple to fermions and can decay to $h^\pm \to Z W^\pm$. This possibility changes the
 direct bounds derived from searches for a charged Higgs at the LHC, as well as the indirect bounds on flavour. These changes are due to the doublet-triplet mixing in the charged Higgs and chargino sectors of the triplet extended model \cite{tripch}. Such sectors can 
 be very useful in order to establish the $SU(2)$ content of the extra scalars, since in this model a very light triplet-like charged Higgs states cannot be ruled out \cite{pbancc}. 
 
 Finally, the superpartners of this triplet- and singlet- like scalars can be dark matter candidates. In particular, a light pseudoscalar sector provides the much needed
 annihilation channel in order to respect the correct dark matter relic density. As we have seen, 
 both direct and indirect constraints can play a significant role in the searches for scalars in higher representations of the $SU(2)$ gauge symmetry, setting a clear distinction respect to the ordinary doublet construction, which is typical of the SM. 
 
 Our approach, though specific to the light pseudoscalar sector of the TNMSSM, can be extended to other models, not necessarily supersymmetric. For instance, it could apply, generically, to scenarios in which the SM Higgs mixes with a  scalar state, for instance a dilaton, as expected in a possible conformal extension of the SM \cite{dil}.  Being the dilaton the pseudo Nambu-Goldstone mode of broken scale invariance, and hence very light, we expect some similarities in the analysis. This is left to future work.

 \vspace{1cm}
 
 \centerline{\bf Acknowledgement} 
The work of C.C. is supported by a {\em The Leverhulme Trust Visiting Professorship}
at the University of Southampton in the STAG Research Centre and Mathematical Sciences Department.
He thanks Kostas Skenderis and the members of the Centre for discussions and the stimulating atmosphere.

\newpage 

\appendix
\section{Higgs coupling to pseudoscalars}
Here we report the vertex $g_{h_i a_j a_k}$ which we used in the calculation of the decay width $\Gamma_{h_1\rightarrow a_1,a_1}$. The vertex is
\begin{align}
g_{h_i a_j a_k}&=i \sqrt{2} A_\kappa \mathcal{R}^P_{j3} \mathcal{R}^P_{k3}
   \mathcal{R}^S_{i3}-\frac{i}{\sqrt{2}} A_S \Big[\mathcal{R}^P_{j3}
   \Big(\mathcal{R}^P_{k1}   	\mathcal{R}^S_{i2}+\mathcal{R}^P_{k2}\mathcal{R}^S_{i1}\Big)+\mathcal{R}^P_{j1} \Big(\mathcal{R}^P_{k3}
   \mathcal{R}^S_{i2}+\mathcal{R}^P_{k2}
   \mathcal{R}^S_{i3}\Big)\nn\\
   &+\mathcal{R}^P_{j2} \Big(\mathcal{R}^P_{k3}
   \mathcal{R}^S_{i1}+\mathcal{R}^P_{k1}
   \mathcal{R}^S_{i3}\Big)\Big]+\frac{i}{2} A_T
   \Big[\mathcal{R}^P_{j4} \Big(\mathcal{R}^P_{k1}
   \mathcal{R}^S_{i2}+\mathcal{R}^P_{k2}
   \mathcal{R}^S_{i1}\Big)+\mathcal{R}^P_{j1} \Big(\mathcal{R}^P_{k4}
   \mathcal{R}^S_{i2}+\mathcal{R}^P_{k2}
   \mathcal{R}^S_{i4}\Big)\nn\\
   &+\mathcal{R}^P_{j2} \Big(\mathcal{R}^P_{k4}
   \mathcal{R}^S_{i1}+\mathcal{R}^P_{k1} \mathcal{R}^S_{i4}\Big)\Big]+i
   \sqrt{2} A_{TS} \Big[\mathcal{R}^P_{j3} \mathcal{R}^P_{k4}
   \mathcal{R}^S_{i4}+\mathcal{R}^P_{j4} \Big(\mathcal{R}^P_{k4}
   \mathcal{R}^S_{i3}+\mathcal{R}^P_{k3}
   \mathcal{R}^S_{i4}\Big)\Big]\nn\\
   &-\frac{i}{2}v_S \Big[2 \lambda_S
    \Big(\Big(\lambda_S  \mathcal{R}^P_{j2}+\kappa  \mathcal{R}^P_{j1}\Big)
   \mathcal{R}^P_{k2}+\Big(\kappa  \mathcal{R}^P_{j2}+\lambda_S 
   \mathcal{R}^P_{j1}\Big) \mathcal{R}^P_{k1}\Big) \mathcal{R}^S_{i3}-2 \kappa  \lambda_S 
   \mathcal{R}^P_{k3} \Big(\mathcal{R}^P_{j1}
   \mathcal{R}^S_{i2}+\mathcal{R}^P_{j2} \mathcal{R}^S_{i1}\Big)\nn\\
   &+\sqrt{2}
   \lambda_T \Big(\lambda_{TS} \mathcal{R}^P_{k4} \Big(\mathcal{R}^P_{j1}
   \mathcal{R}^S_{i2}+\mathcal{R}^P_{j2} \mathcal{R}^S_{i1}\Big)-\Big(\Big(\lambda_S
    \mathcal{R}^P_{j2}+\lambda_{TS} \mathcal{R}^P_{j1}\Big)
   \mathcal{R}^P_{k2}\nn\\
   &+\Big(\lambda_{TS} \mathcal{R}^P_{j2}+\lambda_S 
   \mathcal{R}^P_{j1}\Big) \mathcal{R}^P_{k1}\Big)
   \mathcal{R}^S_{i4}\Big)+\lambda_{TS} \mathcal{R}^P_{j4} \Big(\sqrt{2}
   \lambda_T \Big(\mathcal{R}^P_{k1}
   \mathcal{R}^S_{i2}+\mathcal{R}^P_{k2} \mathcal{R}^S_{i1})+4 \Big(2
   \lambda_{TS}-\kappa \Big) \mathcal{R}^P_{k4} \mathcal{R}^S_{i3}\nn\\
   &+4 \kappa 
   \mathcal{R}^P_{k3} \mathcal{R}^S_{i4}\Big)+2 \kappa 
   \mathcal{R}^P_{j3} \Big(2
   \kappa  \mathcal{R}^P_{k3} \mathcal{R}^S_{i3}+2 \lambda_{TS}
   \mathcal{R}^P_{k4} \mathcal{R}^S_{i4}-\lambda_S  \mathcal{R}^P_{k1}
   \mathcal{R}^S_{i2}-\lambda_S  \mathcal{R}^P_{k2} \mathcal{R}^S_{i1}\Big)\Big]\nn\\
   &-\frac{i}{2}v_T
   \Big[\Big(\mathcal{R}^P_{j2} \mathcal{R}^P_{k2}+\mathcal{R}^P_{j1}
   \mathcal{R}^P_{k1}\Big) \mathcal{R}^S_{i4} \lambda_T^2+\sqrt{2}
   \Big(\lambda_{TS} \mathcal{R}^P_{k3} \Big(\mathcal{R}^P_{j1}
   \mathcal{R}^S_{i2}+\mathcal{R}^P_{j2}
   \mathcal{R}^S_{i1}\Big)\nn\\
   &+\lambda_{TS} \mathcal{R}^P_{j3}
   \Big(\mathcal{R}^P_{k1} \mathcal{R}^S_{i2}+\mathcal{R}^P_{k2}
   \mathcal{R}^S_{i1}\Big)-\Big(\mathcal{R}^P_{j2} (\lambda_S 
   \mathcal{R}^P_{k2}+\lambda_{TS}
   \mathcal{R}^P_{k1})+\mathcal{R}^P_{j1} \Big(\lambda_{TS}
   \mathcal{R}^P_{k2}+\lambda_S  \mathcal{R}^P_{k1}\Big)\Big)
   \mathcal{R}^S_{i3}\Big) \lambda_T\nn\\
   &+2 \lambda_{TS} \Big(2 \kappa
    \Big(\mathcal{R}^P_{j4} \mathcal{R}^P_{k3}+\mathcal{R}^P_{j3}
   \mathcal{R}^P_{k4}\Big) \mathcal{R}^S_{i3}-\lambda_S 
   \Big(\mathcal{R}^P_{k4} \Big(\mathcal{R}^P_{j1}
   \mathcal{R}^S_{i2}+\mathcal{R}^P_{j2}
   \mathcal{R}^S_{i1}\Big)+\mathcal{R}^P_{j4} \Big(\mathcal{R}^P_{k1}
   \mathcal{R}^S_{i2}+\mathcal{R}^P_{k2} \mathcal{R}^S_{i1}\Big)\Big)\nn\\
   &+\Big(\lambda  \mathcal{R}^P_{j1}
   \mathcal{R}^P_{k2}+\lambda_S  \mathcal{R}^P_{j2} \mathcal{R}^P_{k1}+2
   \Big(2 \lambda_{TS}-\kappa \Big) \mathcal{R}^P_{j3} \mathcal{R}^P_{k3}+2
   \lambda_{TS} \mathcal{R}^P_{j4} \mathcal{R}^P_{k4}\Big)
   \mathcal{R}^S_{i4}\Big)\Big]\nn\\
   &-\frac{i}{4}v_u
   \Big[\Big(\left(g_Y^2+g_L^2\right) \mathcal{R}^P_{j1}
   \mathcal{R}^P_{k1}-\left(g_Y^2+g_L^2-2 \lambda_T^2-4 \lambda_S
   ^2\right) \mathcal{R}^P_{j2} \mathcal{R}^P_{k2}\Big)
   \mathcal{R}^S_{i1}\nn\\
   &+2 \mathcal{R}^P_{j4} \Big(\mathcal{R}^P_{k4}
   \Big(\mathcal{R}^S_{i1} \lambda_T^2+2 \lambda_{TS} \lambda_S 
   \mathcal{R}^S_{i2}\Big)-\sqrt{2} \lambda_T \mathcal{R}^P_{k3}
   \Big(\lambda_{TS} \mathcal{R}^S_{i2}+\lambda_S 
   \mathcal{R}^S_{i1}\Big)\nn\\
   &+\lambda_{TS} \mathcal{R}^P_{k2} \left(\sqrt{2}
   \lambda_T \mathcal{R}^S_{i3}-2 \lambda_S 
   \mathcal{R}^S_{i4}\right)\Big)+2 \mathcal{R}^P_{j2}
   \Big(\mathcal{R}^P_{k3} \left(\sqrt{2} \lambda_T \lambda_{TS}
   \mathcal{R}^S_{i4}-2 \kappa  \lambda_S 
   \mathcal{R}^S_{i3}\right)\nn\\
   &+\lambda_{TS} \mathcal{R}^P_{k4} \Big(\sqrt{2}
   \lambda_T \mathcal{R}^S_{i3}-2 \lambda_S 
   \mathcal{R}^S_{i4}\Big)\Big)+2 \mathcal{R}^P_{j3} \Big(2 \lambda_S 
   \mathcal{R}^P_{k3} \Big(\kappa  \mathcal{R}^S_{i2}+\lambda_S 
   \mathcal{R}^S_{i1}\Big)-2 \kappa  \lambda_S  \mathcal{R}^P_{k2}
   \mathcal{R}^S_{i3}\nn\\
   &-\sqrt{2} \lambda_T \Big(\mathcal{R}^P_{k4}
   \Big(\lambda_{TS} \mathcal{R}^S_{i2}+\lambda_S 
   \mathcal{R}^S_{i1}\Big)-\lambda_{TS} \mathcal{R}^P_{k2}
   \mathcal{R}^S_{i4}\Big)\Big)\Big]\nn\\
   &-\frac{i}{4}v_d
   \Big[\left(g_Y^2+g_L^2\right) \mathcal{R}^P_{j2}
   \mathcal{R}^P_{k2} \mathcal{R}^S_{i2}-\mathcal{R}^P_{j1}
   \Big(\left(g_Y^2+g_L^2-2 \lambda_T^2-4 \lambda_S^2\right)
   \mathcal{R}^P_{k1} \mathcal{R}^S_{i2}\nn\\
   &+\mathcal{R}^P_{k3} \left(4
   \kappa  \lambda_S  \mathcal{R}^S_{i3}-2 \sqrt{2} \lambda_T \lambda_{TS}
   \mathcal{R}^S_{i4}\right)-2 \lambda_{TS} \mathcal{R}^P_{k4}
   \left(\sqrt{2} \lambda_T \mathcal{R}^S_{i3}-2 \lambda_S 
   \mathcal{R}^S_{i4}\right)\Big)\nn\\
   &+2 \Big(\mathcal{R}^P_{j4}
   \Big(\mathcal{R}^P_{k4} \Big(\mathcal{R}^S_{i2}
   \lambda_T^2+2 \lambda_{TS} \lambda_S  \mathcal{R}^S_{i1}\Big)-\sqrt{2} \lambda_T \mathcal{R}^P_{k3} \Big(\lambda_S 
   \mathcal{R}^S_{i2}+\lambda_{TS}
   \mathcal{R}^S_{i1}\Big)\nn\\
   &+\lambda_{TS}
   \mathcal{R}^P_{k1} \Big(\sqrt{2} \lambda_T \mathcal{R}^S_{i3}-2
   \lambda_S  \mathcal{R}^S_{i4}\Big)\Big)+\mathcal{R}^P_{j3} \Big(2
   \lambda_S  \mathcal{R}^P_{k3} \Big(\lambda_S  \mathcal{R}^S_{i2}+\kappa 
   \mathcal{R}^S_{i1}\Big)-2 \kappa  \lambda_S  \mathcal{R}^P_{k1}
   \mathcal{R}^S_{i3}\nn\\
   &-\sqrt{2} \lambda_T \Big(\mathcal{R}^P_{k4} \Big(\lambda_S 
   \mathcal{R}^S_{i2}+\lambda_{TS} \mathcal{R}^S_{i1}\Big)-\lambda_{TS}
   \mathcal{R}^P_{k1} \mathcal{R}^S_{i4}\Big)\Big)\Big)\Big]
\end{align}
%%%%%%%%%%%%%%%%%%%%%%%%%%%%%%%%%
%%%%%%%%%%%%%%%%%%%%%%%%%%%%%%%%%

\end{document}